\documentclass{emulateapj}
\usepackage{apjfonts}

\newcommand{\etal}{et~al.}
\newcommand{\MgIIdblt}{{\rm Mg}\kern 0.1em{\sc ii}~$\lambda\lambda 2796, 2803$}
\newcommand{\CaII}{\hbox{{\rm Ca}\kern 0.1em{\sc ii}}}
\newcommand{\CIII}{\hbox{{\rm C}\kern 0.1em{\sc iii}}}
\newcommand{\CIV}{\hbox{{\rm C}\kern 0.1em{\sc iv}}}
\newcommand{\HI}{\hbox{{\rm H}\kern 0.1em{\sc i}}}
\newcommand{\HII}{\hbox{{\rm H}\kern 0.1em{\sc ii}}}
\newcommand{\Lya}{\hbox{{\rm Ly}\kern 0.1em$\alpha$}}
\newcommand{\Lyb}{\hbox{{\rm Ly}\kern 0.1em$\beta$}}
\newcommand{\MgI}{\hbox{{\rm Mg}\kern 0.1em{\sc i}}}
\newcommand{\MgII}{\hbox{{\rm Mg}\kern 0.1em{\sc ii}}}
\newcommand{\MgIII}{\hbox{{\rm Mg}\kern 0.1em{\sc iii}}}
\newcommand{\NI}{\hbox{{\rm N}\kern 0.1em{\sc i}}}
\newcommand{\NII}{\hbox{{\rm N}\kern 0.1em{\sc ii}}}
\newcommand{\NV}{\hbox{{\rm N}\kern 0.1em{\sc v}}}
\newcommand{\OVI}{\hbox{{\rm O}\kern 0.1em{\sc vi}}}
\newcommand{\OI}{\hbox{{\rm O}\kern 0.1em{\sc i}}}
\newcommand{\OII}{\hbox{[{\rm O}\kern 0.1em{\sc ii}]}}
\newcommand{\OIV}{\hbox{{\rm O}\kern 0.1em{\sc iv}]}}
\newcommand{\kms}{\hbox{km~s$^{-1}$}}

\slugcomment{Accepted to ApJ: March 9, 2007}

\shorttitle{\sc {\MgII} Absorber Galaxy Morphology}
\shortauthors{\sc Kacprzak {\etal}}

\begin{document}

\title{A Correlation Between Galaxy Morphology and {\MgII} Halo
Absorption Strength}

\author{\sc
Glenn G. Kacprzak\altaffilmark{1}, 
Christopher W. Churchill\altaffilmark{1},  
Charles C. Steidel\altaffilmark{2}, \\
Michael T. Murphy\altaffilmark{3},  
and
Jessica L. Evans\altaffilmark{1}
}

\altaffiltext{1}{New Mexico State University, Las Cruces, NM 88003
{glennk@nmsu.edu, cwc@nmsu.edu, jlevans@nmsu.edu}}

\altaffiltext{2}{Caltech, Pasadena, CA 91125 
{ccs@astro.caltech.edu}}

\altaffiltext{3}{Institute of Astronomy, Cambridge CB3 0HA, UK 
{mim@ast.cam.ac.uk}}

\begin{abstract}

We compared the quantified morphological properties of 37 intermediate
redshift ($0.3 \leq z \leq 1$) {\MgII} absorption selected galaxies to
the properties of the absorbing halo gas [$0.03 \leq W_r(2796) \leq
2.90$~{\AA}], observed in the spectra of background quasars.  The
galaxy morphologies were measured using GIM2D modeling of {\it Hubble
Space Telescope\/} WFPC--2 images and the absorbing gas properties
were obtained from HIRES/Keck and UVES/VLT quasar spectra.  We found a
$3.1~\sigma$ correlation between galaxy morphological asymmetries
normalized by the quasar--galaxy projected separations, $A/D$, and the
{\MgII} rest--frame equivalent widths.  Saturation effects cause
increased scatter in the relationship with increasing $W_r(2796)$.  We
defined a subsample for which the fraction of saturated pixels in the
absorption profiles is $f_{sat} < 0.5$.  This criterion resulted in a
subsample of 28 systems with $W_r(2796) \leq 1.39$~{\AA}.  The
correlation strengthened to $3.3~\sigma$.  We also find a paucity of
small morphological asymmetries for galaxies selected by {\MgII}
absorption as compared to those of the general population of field
galaxies, as measured in the Medium Deep Survey. The K--S probability
that the two samples are drawn from the same galaxy population is
ruled out at a 99.8\% confidence level.  Based upon four different
measures of galaxy asymmetry, it is evident that the morphological
perturbations of galaxies selected by {\MgII} absorption are ``minor''
and centrally concentrated.  The $A/D$--$W_r(2796)$ correlation
suggests a connection between the processes that perturb galaxies and
the quantity of gas in their halos, normalized by the impact
parameter.  Since the perturbations are minor, it is clear that
dramatic processes or events are not required for a galaxy to have an
extended halo; the galaxies appear ``normal''.  We suggest that
common, more mild processes that populate halos with gas, such as
satellite galaxy merging, accretion of the local cosmic web, and
longer--range galaxy--galaxy interactions, consequently also induce
the observed minor perturbations in the galaxies.

\end{abstract}



\section{Introduction}
\label{sec:introduction}

Metal--line absorbing galaxies are selected by their gas cross
sections, i.e., the presence of extended, associated gaseous halos
detected in quasar absorption lines. Since the kinematic, chemical,
and ionization conditions of the gaseous halos can be studied in
detail and compared directly to the galaxies themselves, these
galaxy--absorber pairs provide insight into the role of gas in galaxy
evolution. A connection between galaxy morphological characteristics
and absorption properties would provide constraints on competing
scenarios of the dynamics, enrichment, and geometry of halos.

The use of high spatial resolution imaging, such as WFPC--2/{\it
HST\/} and ACS/{\it HST}, provides the opportunity to study the
morphology of intermediate redshift galaxies with unprecedented
detail. One of the first techniques for classifying and quantifying
the morphologies of intermediate--to--high redshift galaxies is the
$C$--$A$ method \citep{abraham94}, where $C$ is the concentration of
light and $A$ is the asymmetry of the galaxies. This technique has
been applied to magnitude--limited samples in the Hubble Deep Field
\citep{abraham96a}, the Medium Deep Survey \citep{abraham96b}, and the
Ohio State University Bright Spiral Galaxy Survey \citep{whyte02}.

\citet{simard02} developed GIM2D, a two--dimensional decomposition
fitting program that models structural parameters of galaxies. This
program has been used to quantify the morphologies of Hubble Deep
Field North galaxies \citep{marleau98}, the luminosity--size relation
of field disk galaxies for $0.1 \leq z \leq 1.1$ \citep{simard99}, the
number density and luminosity function of E/S0 galaxies for $z \leq 1$
\citep{im02}, and the morphologies and bar structures of field
galaxies at $0.4 \leq z \leq 1.1$ \citep{zheng05}.

Despite the availability of these tools, there has been little effort
put forth toward quantifying the morphology of galaxies selected by
metal--line absorption in quasar spectra. To our knowledge, there is
only one study in which the morphologies of two {\OVI} absorbing
galaxies were quantified using the $C$--$A$ method \citep{savage98}.
These galaxies have impact parameters $100~h_{75}^{-1}$ and
$350~h_{75}^{-1}$ kpc, respectively ($q_0 = 0.05$, $\Omega _{\Lambda}
= 0$).  Based upon the concentration parameter, they find that these
two galaxies are Sbc--Sc spirals.  They also find that their
asymmetries are lower than that of strongly interacting or irregular
galaxies.

Though various classes of metal--line absorbers are available for
sample selection, the {\MgIIdblt} doublet is arguably the best for
studying the association of absorption with galaxies.  {\MgII} is
observable from the ground for intermediate redshifts, a regime where
galaxies can be well resolved in {\it HST\/} images.  {\MgII}
absorption probes low ionization halo and is detectable in
environments spanning five decades of {\HI} column densities
\citep{archiveI}, from sub--Lyman limit \citep[e.g.,][]{weakI} to
damped {\Lya} systems \citep[e.g.,][]{rao00}. The association of
normal, bright, field galaxies with {\MgII} absorption at moderate
impact parameters has been well established
\citep[e.g.,][]{bb91,sdp94}. 

Much of our current understanding of {\MgII} absorption selected
galaxies is based upon the galaxy $B$ and $K$ luminosities,
rest--frame $B-K$ colors, and impact parameters
\citep{sdp94,csv96,archiveII}.  The spectroscopic types of these
galaxies have been based upon their colors from ground--based images
\citep[e.g.,][]{bb91,sdp94,chen03}, or, in a few cases, based upon
their spectra \citep[e.g.,][]{spinrad93,gb97}. \citet{sdp94}
determined that the average $B-K$ color for {\MgII} absorbing galaxies
is consistent with that of an Sb galaxy, though all galaxy colors
redder than an Sd color are represented.  Using Sloan $g$, $r$, and
$i$ band images toward 700 quasars, \citet{zibetti05,zibetti06} also
find galaxy colors similar to those of local Sb--c spirals based upon
statistical methods.  Space--based WFPC--2/{\it HST\/} images of
intermediate redshift absorption selected galaxies qualitatively
reveal that their morphological types appear to be similar to those of
local spiral and elliptical galaxies
\citep{steidel98,chen01,chen03,glenn-china,cwc-china}.

In the local universe, where case--by--case studies can be very
detailed, the morphologies of galaxies have provided important
insights for understanding the characteristics of gaseous halos.
There are several case studies where faint halo gas is observed in
emission to a projected distance of 13~kpc; the majority of these
galaxies have either been harassed by and/or are interacting with
other galaxies, as is apparent from their morphological asymmetries
and {\HI} warps \citep{puche92,swaters97,rand00}. If local interacting
or perturbed galaxies have more gas ejected into their halos, then
intermediate redshift metal--line absorption selected galaxies might
likewise display some evidence of morphological perturbations. If so,
what can a galaxy's morphology and/or morphological perturbations tell
us about the absorption characteristics of the halo gas? Moreover,
what can galaxy perturbations tell us about the mechanisms that
populate galaxy halos with gas?

In order to shed light on these questions, we have applied GIM2D on
WFPC--2/{\it HST\/} images to quantify the morphologies of known
{\MgII} absorbing galaxies at intermediate redshifts. We aim to
examine whether the morphologies of metal--line absorption selected
galaxies can provide insight into the quantity and distribution of gas
residing in their extended halos.  We have employed several asymmetry
measures, since each invokes different methods for quantifying
deviations from an axisymmetric, smooth morphology. Our study has
revealed a correlation between the rest--frame equivalent width of the
{\MgII} $\lambda 2796$ transition, $W_r(2796)$, and the galaxy
morphological asymmetries normalized by the impact parameter, i.e.,
the projected galactocentric distance to the absorption.

This paper is organized as follows: In \S~\ref{sec:data}, we present
our sample selection methodology, the resulting sample, and the data
reduction methods we employed.  In \S~\ref{sec:modeling}, the basic
details of quantifying the galaxy morphological asymmetries are
presented.  We present our results and analyze the differences between
various methods of measuring morphological asymmetries in
\S~\ref{sec:results}.  In \S~\ref{sec:discussion}, we discuss the
implications of our analysis.  Concluding remarks are provided in
\S~\ref{sec:conclusion}.  Throughout this paper, we adopt a
$H_0=70$~{\kms} Mpc$^{-1}$, $\Omega_m=0.3$, and $\Omega_{\Lambda}=0.7$
cosmology.

\section{Galaxy Selection and Data Analysis}
\label{sec:data}

Our sample of galaxies is selected by the presence of {\MgII}
absorption in ground--based, optical quasar spectra, primarily high
resolution ($R = 45,000, \hbox{\rm FWHM} \simeq 6$~{\kms}) HIRES/Keck
\citep{vogt94} or UVES/VLT \citep{dekker00} data.  These high
signal--to--noise ratio spectra provide a {\MgII} $\lambda 2796$
rest--frame equivalent width detection threshold of $W_r(2796) \simeq
0.02$~{\AA}.  Thus, we have been able to select galaxies associated
with a wide range of $W_r(2796)$, including the so called ``weak''
systems, i.e., $W_r(2796) < 0.3$~{\AA} \citep{weakI,cwcweakgals}.

In the cases where we have a HIRES or UVES quasar spectrum of an
{\MgII} absorber, we have measured the absorption properties and
equivalent widths from the data.  The HIRES spectra were reduced using
IRAF\footnote{IRAF is written and supported by the IRAF programming
group at the National Optical Astronomy Observatories (NOAO) in
Tucson, Arizona. NOAO is operated by the Association of Universities
for Research in Astronomy (AURA), Inc.\ under cooperative agreement
with the National Science Foundation.}. The UVES spectra were reduced
using the standard ESO pipeline and a custom code called the UVES
Post--Pipeline Echelle Reduction \citep[{\sc uves popler},][]{popler}.
Analysis of the {\MgII} absorption profiles was performed using
graphic--based interactive software of our own design
\citep{weakI,archiveI,cv01,cvc03} for local continuum fitting,
objective feature identification, and for measuring absorption
properties. For {\MgII} absorbers where we do not have a HIRES or UVES
quasar spectrum, we obtained the measured $W_r(2796)$ from the
literature.

\begin{deluxetable*}{llcrcrcccc}
\tabletypesize{\scriptsize}
\tablecaption{Galaxy--Absorber Sample\label{tab:sample}}
\tablecolumns{10}
\tablewidth{0pt}
\tablehead{
\colhead{QSO Field}&
\colhead{$z_{abs}$} &
\colhead{$W_r(2796)$} &
\colhead{{\MgII}\tablenotemark{a}} &
\colhead{$D$ (kpc)} &
\colhead{ID$_{gal}$\tablenotemark{b}} &
\colhead{$m_{HST}$} &
\colhead{Filter} &
\colhead{Exp, sec} &
\colhead{PID}
}
\startdata
Q$0002+051 $ & $ 0.298059 $ & $ 0.246\pm0.004$       & 1     & $ 59.3\pm0.3$      & 1   & $19.96\pm0.01$ & F702W & 4600 & Steidel/5984  \\
Q$0002+051 $ & $ 0.591365 $ & $ 0.102\pm0.002$       & 1     & $ 36.2\pm0.4$      & 1   & $21.24\pm0.01$ & F702W & 4600 & Steidel/5984  \\
Q$0002+051 $ & $ 0.851407 $ & $ 1.119\pm0.013$       & 1     & $ 25.9\pm0.5$      & 1   & $22.34\pm0.02$ & F702W & 4600 & Steidel/5984  \\
Q$0109+200 $ & $ 0.5346 $ & $ 2.26\phantom{0\pm0.000}$ & 2 & $ 45.1\pm0.4$      & 2,3 & $22.27\pm0.02$ & F702W & 1800 & Disney/6303   \\
Q$0117+213 $ & $ 0.729075 $ & $ 0.244\pm0.006$       & 1     & $ 55.5\pm1.1$      & 1   & $21.06\pm0.02$ & F702W & 2008 & Zuo/6115  \\
Q$0150-202 $ & $ 0.3887 $ & $ 0.58\phantom{0\pm0.000}$ & 2 & $ 60.0\pm0.7$      & 2,3 & $21.14\pm0.01$ & F702W & 5100 & Steidel/6557  \\
Q$0229+131 $ & $ 0.417337 $ & $ 0.816\pm0.022$       & 1     & $ 37.5\pm0.5$      & 2,3 & $19.72\pm0.01$ & F702W & 5000 & Steidel/6557  \\
Q$0235+164 $ & $ 0.5240 $ & $ 2.34\phantom{0}\pm0.05\phantom{0}$ & 3     & $ 12.1\pm0.6$  & 3   & $20.30\pm0.02$ & F702W & 600  & Burbidge/5096 \\
Q$0450-132 $ & $ 0.493937 $ & $ 0.674\pm0.026$       & 1     & $ 50.1\pm0.4$      & 1   & $21.55\pm0.01$ & F702W & 2500 & Steidel/5984  \\
Q$0450-132 $ & $ 0.493937 $ & $ 0.674\pm0.026$       & 1     & $ 62.7\pm0.7$      & 1   & $21.53\pm0.01$ & F702W & 2500 & Steidel/5984  \\
Q$0827+243 $ & $ 0.524968 $ & $ 2.419\pm0.017$       & 1     & $ 37.5\pm0.9$      & 4   & $20.64\pm0.01$ & F702W & 4600 & Steidel/5984  \\
Q$0836+113 $ & $ 0.7874 $ & $ 2.71\phantom{0\pm0.000}$ & 4     & $ 26.9\pm0.9$  & 5   & $22.63\pm0.02$ & F702W & 5000 & Steidel/6557  \\
Q$1019+309 $ & $ 0.3461 $ & $ 0.70\phantom{0}\pm0.07\phantom{0}$ & 5 & $46.0\pm0.5$  & 1   & $20.47\pm0.01$ & F702W & 5100 & Steidel/6557  \\
Q$1038+064 $ & $ 0.441453 $ & $ 0.673\pm0.014$       & 1     & $ 56.0\pm0.6$      & 6   & $20.70\pm0.01$ & F702W & 4600 & Steidel/5984  \\
Q$1127-145 $ & $ 0.312710 $ & $ 1.773\pm0.006$       & 1     & $ 81.0\pm0.3$      & 2,3 & $19.67\pm0.01$ & F814W & 4400 & Bechtold/9173 \\
Q$1127-145 $ & $ 0.312710 $ & $ 1.773\pm0.006$       & 1     & $ 45.6\pm0.3$      & 2,3 & $19.48\pm0.01$ & F814W & 4400 & Bechtold/9173 \\
Q$1148+387 $ & $ 0.553362 $ & $ 0.644\pm0.014$       & 1     & $ 20.3\pm0.6$      & 6   & $20.94\pm0.01$ & F702W & 4800 & Steidel/5984  \\
Q$1209+107 $ & $ 0.392925 $ & $ 1.210\pm0.006$       & 1     & $ 37.9\pm0.4$      & 3   & $21.74\pm0.02$ & F702W & 3600 & Bergeron/5351 \\
Q$1222+228 $ & $ 0.550196 $ & $ 0.098\pm0.009$       & 1     & $ 36.4\pm1.1$      & 6   & $22.50\pm0.02$ & F702W & 5000 & Steidel/5984  \\ 
Q$1241+176 $ & $ 0.550483 $ & $ 0.482\pm0.013$       & 1     & $ 21.4\pm0.4$      & 1   & $21.40\pm0.01$ & F702W & 5000 & Steidel/6557  \\ 
Q$1246-057 $ & $ 0.639908 $ & $ 0.461\pm0.006$       & 1     & $ 29.9\pm0.8$      & 1   & $22.21\pm0.02$ & F702W & 4600 & Steidel/5984  \\
Q$1317+277 $ & $ 0.660053 $ & $ 0.348\pm0.007$       & 1     & $103.2\pm0.6$      & 6   & $21.35\pm0.01$ & F702W & 4700 & Steidel/5984  \\
Q$1332+552 $ & $ 0.3740 $ & $ 2.90\phantom{0\pm0.000}$ & 2 & $ 27.9\pm0.5$      & 7   & $19.40\pm0.01$ & F702W & 2800 & Steidel/6557  \\
Q$1511+103 $ & $ 0.4369 $ & $ 0.454\pm0.046$       & 6     & $ 38.2\pm0.4$      & 2,3 & $21.22\pm0.01$ & F702W & 5000 & Steidel/6557  \\
Q$1622+235 $ & $ 0.317596 $ & $ 0.491\pm0.012$       & 1     & $ 54.6\pm0.5$      & 8   & $19.45\pm0.01$ & F702W & 24,000 & Steidel/5304 \\
Q$1622+235 $ & $ 0.368113 $ & $ 0.253\pm0.008$       & 1     & $114.0\pm0.4$      & 8   & $20.00\pm0.01$ & F702W & 24,000 & Steidel/5304 \\
Q$1622+235 $ & $ 0.471930 $ & $ 0.769\pm0.009$       & 1     & $ 34.1\pm0.5$      & 8   & $22.27\pm0.01$ & F702W & 24,000 & Steidel/5304 \\
Q$1622+235 $ & $ 0.656103 $ & $ 1.451\pm0.011$       & 1     & $ 99.6\pm0.5$      & 8   & $22.52\pm0.02$ & F702W & 24,000 & Steidel/5304 \\
Q$1622+235 $ & $ 0.702903 $ & $ 0.032\pm0.003$       & 1     & $112.7\pm0.9$      & 8   & $21.63\pm0.01$ & F702W & 24,000 & Steidel/5304 \\
Q$1622+235 $ & $ 0.797079 $ & $ 0.468\pm0.009$       & 1     & $ 71.5\pm1.3$      & 8   & $22.37\pm0.02$ & F702W & 24,000 & Steidel/5304 \\
Q$1622+235 $ & $ 0.891280 $ & $ 1.548\pm0.008$       & 1     & $ 23.2\pm0.5$      & 8   & $22.65\pm0.02$ & F702W & 24,000 & Steidel/5304 \\
Q$1623+268 $ & $ 0.8885 $ & $ 0.68\phantom{0}\pm0.03\phantom{0}$ & 7 & $48.2\pm0.5$ & 1   & $23.64\pm0.03$ & F702W & 4600 & Steidel/5984  \\
Q$1623+268 $ & $ 0.8875 $ & $ 0.27\phantom{0}\pm0.02\phantom{0}$ & 7 & $69.9\pm1.3$ & 1   & $23.80\pm0.06$ & F702W & 4600 & Steidel/5984  \\
Q$2128-123 $ & $ 0.429820 $ & $ 0.464\pm0.004$       & 1     & $ 48.8\pm0.5$      & 2,3 & $20.42\pm0.01$ & F702W & 1800 & Mecchetto/5143 \\
Q$2206-199 $ & $ 0.751923 $ & $ 0.886\pm0.003$       & 1     & $ 44.2\pm0.7$      & 3   & $22.35\pm0.01$ & F702W & 5000 & Steidel/6557  \\
Q$2206-199 $ & $ 0.948361 $ & $ 0.253\pm0.002$       & 1     & $ 87.2\pm0.5$      & 3   & $21.92\pm0.01$ & F702W & 5000 & Steidel/6557  \\
Q$2206-199 $ & $ 1.017040 $ & $ 1.057\pm0.005$       & 1     & $104.6\pm1.4$      & 3   & $20.99\pm0.01$ & F702W & 5000 & Steidel/6557  \\
\enddata
\tablenotetext{a}{{\MgII} Absorption: (1) This paper, 
 (2)~\citet{gb97}, (3)~\citet{lanzetta92}, (4)~\citet{sdp94},
 (5)~\citet{steidel92}, (6) \citet{foltz86}, 
 (7)~\citet{sargent88}.}
\tablenotetext{b}{Galaxy Identification: (1)~\citet{sdp94},
(2)~\citet{bb91}, (3)~\citet{gb97}, (4)~\citet{kanekar01},
(5)~\citet{lowenthal90}, (6)~\citet{steidel02}, and
(7)~\citet{miller87}, (8)~\citet{steidel97}.}
\end{deluxetable*}

We have limited our selection of galaxies to those with
spectroscopically measured redshifts that are consistent with the
redshift of the {\MgII} absorption in the range $0.3 \leq z \leq 1.0$.
The galaxy redshifts are either obtained through our previous work
\citep{sdp94,csv96,steidel97,steidel02}, or from the literature.  In
order to examine the galaxy morphologies, we further limited
our sample to include only those galaxy--absorber pairs in quasar
fields that have been imaged with either the F702W and/or F814W filter
using WFPC--2/{\it HST}.  The WFPC--2 images were obtained
for various research programs and are available in the {\it HST\/}
archive.  The F702W filter provides a broad--band image similar to a
rest--frame Johnson $B$--band filter for galaxies at $z\sim 0.6$. The
F814W filter provides an image similar to the rest--frame $B$--band
for galaxies at $z\sim 0.85$.

All WFPC--2/{\it HST\/} images were reduced using the WFPC--2
Associations Science Products Pipeline (WASPP\footnote{Developed by
the Canadian Astronomy Data Centre (CADC) and the Space
Telescope--European Coordinating Facility (ST--ECF): {\it
http://archive.stsci.edu/hst/wfpc2/pipeline.html}}).  The WFPC--2
astrometry is calibrated to the USNO2 Catalog \citep{usno2}.  Offsets
for combining images are performed either by cross--correlation,
jitter information, and/or image world coordinate system information.
Robust average images are obtained using the method of artificial
skepticism \citep{stetson89}, which uses a continuous weighting scheme
derived directly from the data.  WASPP data quality verifications
include photometric and astrometric accuracy and correctly set
zero--points.  Galaxy photometry was performed using the Source
Extractor (Sextractor) package \citep{bertin96} with a detection
criterion of 1.5~$\sigma$ above background.  The $m_{F702W}$ and
$m_{F814W}$ magnitudes were measured using the WFPC--2 zero points
\citep{whitmore95}, based upon the Vega system.  Uncertainties in
magnitudes range from 0.01 to 0.02. 

Our final sample comprises 37 galaxies with magnitudes in the range
$19.4 \leq m_{HST} \leq 23.8$ selected by ``weak'' \citep{weakI},
``classic'' \citep{archiveII}, and ``strong'' \citep{bond01b} {\MgII}
absorption spanning the equivalent width range $0.03 \leq W_r(2796)
\leq 2.90$~{\AA}.  Note that for three of the absorbers, each have two
galaxies at the same redshifts.  Since it is always possible that a
pair of galaxies can give rise to the absorption and such conditions
can provide further insights into the galaxy--absorption connection,
we do not exclude these pairs from our sample. 

The sample is presented in Table~\ref{tab:sample}, where we list (1)
the quasar field, (2) the {\MgII} absorption redshift, (3) the
rest--frame {\MgII} $\lambda 2796$ equivalent width, $W_r(2796)$, and
uncertainty (when available), (4) the source of quoted {\MgII}
measurements, (5) the quasar--galaxy impact parameter, $D$, and
uncertainty, and (6) the reference(s) for the galaxy identification.
We also list (7) the apparent magnitude of the galaxy and uncertainty
as measured in the WFPC--2 filter (8), (9) the exposure time, and (10)
the PI name and program identification of the WFPC--2 image.  Further
details about the quasar fields, the absorbing galaxies, and the
absorption properties will be published in an upcoming paper
\citep{glenn-thesis}.

The impact parameters, $D$, presented in Table~\ref{tab:sample}, are
computed using the galaxy $1.5~\sigma$ isophotal centroids determined
by Sextractor.  There is a $\sim 0.05$ pixel uncertainty in the
position of the quasar based upon centroiding errors of unresolved
sources in the images.  As described below, we modeled the galaxies
using GIM2D \citep{simard02}, which provides a second isophotal
centroid.  The dominant uncertainty in $D$ is derived from the pixel
offset of the galaxy isophotal center obtained using Sextractor and
the isophotal center of the galaxy model determined by GIM2D.  This
offset is typically about 0.25 pixels.

\begin{figure*}
\figurenum{1$a$}
\plotone{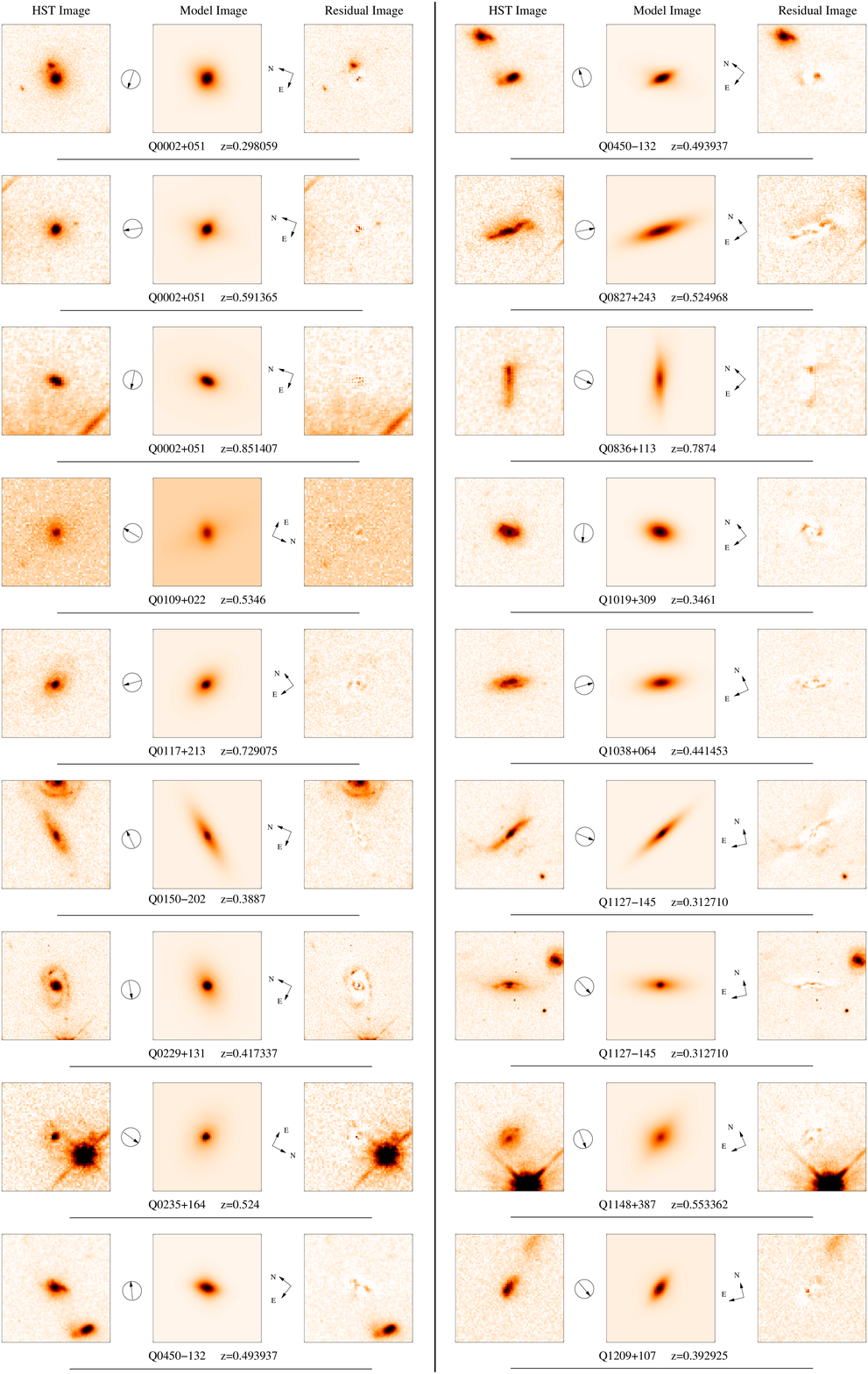}
\caption{(left) WFPC--2 images of galaxies selected by
{\MgII} absorption. The images are 10 times larger than the
$1.5~\sigma$ isophotal area.  --- (center) The GIM2D models of the
galaxies, which provide quantified morphological parameters.  ---
(right) The residual images from the models, showing the underlying
structure and morphological perturbations of the galaxies, which are
quantified using various asymmetry measures.  The encircled arrow
provides the direction to the quasar (galaxy--quasar orientation) and
the cardinal directions are shown. The quasar name and redshift of
{\MgII} absorption is shown under each set of galaxy WFPC--2, model
and residual image. }
\label{fig:gal1}
\end{figure*}

\begin{figure*}
\figurenum{1$b$}
\plotone{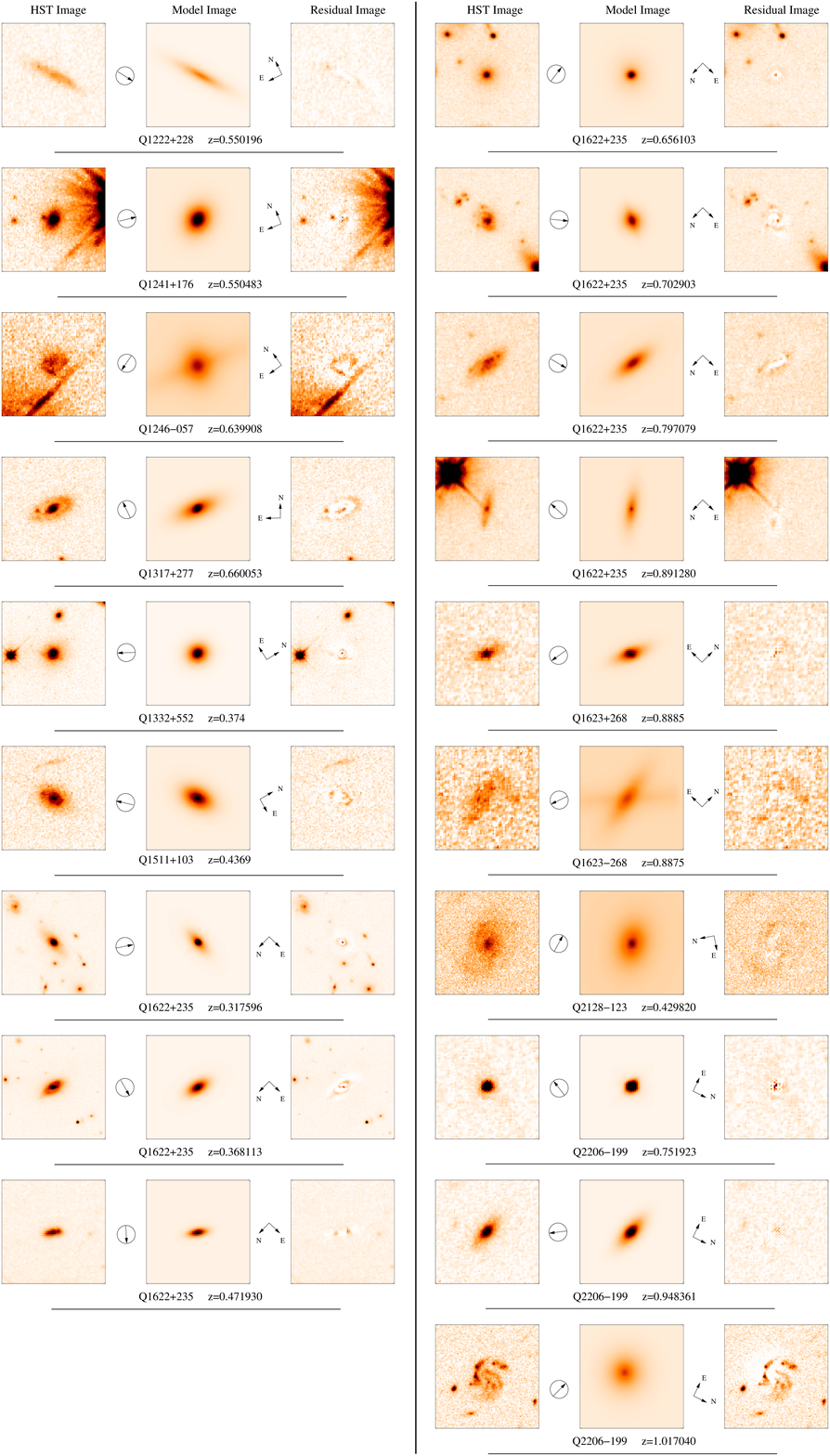}
\caption{ Same as Figure~\ref{fig:gal1}.}
\label{fig:gal2}
\end{figure*}

WFPC--2 portraits of the galaxies are presented in the left--hand
panels of Figure~\ref{fig:gal1}. The portraits have an area 10 times
larger than the $1.5~\sigma$ galaxy isophotal area. The encircled
arrow next to the portrait provides the direction to the quasar
relative to the galaxy; also shown are the cardinal directions.  The
accompanying panels are the image of the GIM2D model and the residual
image.  These are discussed below.

\section{Quantifying Galaxy Asymmetries}
\label{sec:modeling}

For each galaxy, the morphology was quantified by fitting a
two--component (bulge+disk) parametric model to its two--dimensional
surface brightness distribution using GIM2D \citep{simard02}.  We
fitted the surface brightness of the bulge component with a de
Vaucouleurs profile and the surface brightness of the disk
component with an exponential profile.  The centroids of the bulge and
disk components are forced to be co--spatial; they are also fitted
parameters.  We employed this standard profile on all galaxies in our
sample in order to ensure uniform parameterization.  The models are
convolved with the WFPC--2 point spread function, which was determined
at the appropriate locations on the image using Tiny Tim
\citep{krist04}.

Our goal for employing GIM2D to fit {\it HST\/} galaxy images is to
derive structural and morphological parameters that can be used for
quantitative comparison with quantified absorption properties,
primarily the {\MgII} rest--frame equivalent width.  We are
particularly interested in quantified measurements of the
morphological perturbations, or asymmetries in our absorption selected
galaxies.  GIM2D computes four measures of asymmetry using the
formalism of \citet{abraham94}, \citet{schade95}, and
\citet{simard02}.  Portraits of the GIM2D model galaxies (center
panels) and the residual images (right--hand panels) are presented in
Figure~\ref{fig:gal1}.  The residual image is
obtained by subtraction of the GIM2D model image from the {\it HST\/}
image.  For the residuals, regions that lie outside the Sextractor
segmentation image are identical to the {\it HST\/} images.

The first asymmetry measure is obtained from the $C$--$A$ method of
\citet{abraham94}. The Abraham $A$ is computed directly from the data
(model independent) from
\begin{equation}
A\mbox{ }=\mbox{ }\frac{\displaystyle\sum_{ij} \frac{1}{2}\mid
I_{ij}-I_{ij}^{180}\mid}{\displaystyle\sum_{ij} I_{ij}}
-\frac{\displaystyle\sum_{ij} \frac{1}{2}\mid
B_{ij}-B_{ij}^{180}\mid}{\displaystyle\sum_{ij} I_{ij}},
\label{eq:abraham}
\end{equation}
where the $I_{ij}$ are the counts at pixel $ij$ and the $I_{ij}^{180}$
are the counts at the $ij$ location following a 180$^\circ$ rotation
of the galaxy.  The domain of the sums is an elliptical isophotal area
defined by galaxy counts 2~$\sigma$ above background.  The background
counts, $B_{ij}$, are randomly selected from pixels just beyond the
Sextractor segmentation image.  Uncertainties in the $A$ are
determined by computing $A$ for different random selections of the
$B_{ij}$.  In GIM2D, four realizations of background are computed.
For the best value of $A$, we take the average of the four
realizations and for the uncertainty in $A$ we compute the standard
deviation of these realizations about the average.

The second asymmetry measure is the quantity $R_A$ \citep{schade95},
which is also computed using Eq.~\ref{eq:abraham}, except that it
employs the GIM2D galaxy model residuals, $R_{ij}$ and $R_{ij}^{180}$,
instead of the direct counts $I_{ij}$ and $I_{ij}^{180}$ from the
data.  The domain of the sums is also model dependent.  $R_{A}$ is
evaluated for each of the GIM2D galaxy model circular apertures,
1--10~$r_{h}$ (half--light radii) for $R_{ij}$ that are 1, 3 and
5~$\sigma$ above background.  Thus, there are 30 values of $R_A$
representing 10 circular apertures in units of $r_h$ for three
significance thresholds.  As with the Abraham $A$, the background
counts are randomly selected from the pixels just beyond the
Sextractor segmentation image.

Two additional methods, denoted as $A_z$ and $D_z$ \citep{simard02},
are unique to GIM2D. Both are computed directly from the data, but
using the GIM2D model circular apertures.  $A_z$ is evaluated for each
of ten circular apertures, 1--10~$r_{h}$, and is the sum of all
$I_{ij}$ having signal $2$, $3$, and $5~\sigma$ greater than that of
the $180^{\circ}$ symmetrical counterpart normalized by the total
counts.  Similar to the $R_A$, there are 30 values of $A_z$
representing 10 circular apertures in units of $r_h$ for three
significance thresholds.  $D_z$ is computed exclusively {\it
outside\/} the GIM2D model circular aperture 1~$r_h$. It is the sum of
all $I_{ij}$ that are 1.5~$\sigma$ above background for which the
symmetric counterpart $I_{ij}^{180}$ are less than 1.5~$\sigma$ above
background. The sum is normalized by the total galaxy total flux.
$D_z$ is a sensitive measure of highly asymmetric spiral arms, tidal
tails, and highly asymmetric perturbations, since such features tend
to be located to one side of a galaxy (see, for example, Q$1246-057$
and Q$2206-199$ at $z=1.0170$ in Figure~\ref{fig:gal1}). Comparing
$D_z$ to the other asymmetric measures provides insight for
discriminating if a galaxy has suffered strong interactions or only
slight morphological perturbations.

The WFPC--2 image exposure times range from 600~s to 24,000~s, with a
typical time of 5000~s. The quantified morphological parameters
obtained from the 600~s exposure (Q0235+164) should be viewed with
caution. It may also be of concern that the longest exposure of
24,000~s (Q1622+235) could have different measured morphological
parameters if the exposure time was of the order of 5000 seconds. To
verify that the these values did not change drastically as a function
of exposure time, we degraded the 24,000~s exposure to both 10,000 and
5000~s and measured the morphological parameters for these three
exposure times. For all seven galaxies in the degraded 10,000~s
exposure, the measured values are consistent within the uncertainties
of the measurement. For the degraded image of 5000~s, the average
percent difference is 9\% but this is not systematic (some increase
slightly, some decrease slightly). The quoted errors in the
morphological parameters for the Q1622+235 field reflect the errors in
the 5000~s exposure.

\subsection{Effects of PSF Subtraction}
\label{sec:psf}

Seven galaxy--absorber pairs lie at relatively small impact parameters
such that the quasar light potentially contaminates the
two--dimensional light profiles of the galaxies (e.g., Q0002+051
$z=0.851407$, Q0235+164, Q0827+243, Q1148+387, Q1241+176, Q1246-057,
and Q1622+235 $z=0.891280$). In some of these cases, a diffraction
spike passes through the galaxy or passes near the galaxy.  In other
cases, the galaxy lies within the extended tails of the complicated
point spread function (PSF) illumination pattern of the quasar.
Whereas galaxies far from the quasar light pattern are observed to
have a uniform (flat) background, those near the quasar light pattern
lie on a tilted and perhaps slightly non--uniform ``background''.  In
some cases, there is a ramping of the background of roughly 3--5
counts over a 60 pixel radial cut toward the quasar in the vicinity of
the galaxy.

In both cases, the quasar light can affect the physical boundaries of
the Sextractor segmentation image and/or affect the measured
morphological parameters obtained by GIM2D.  We examined the effects
of the quasar light contamination by performing two tests.  First,
when possible, we modeled and subtracted the PSF of the quasar from
the image and compared the resulting segmentation images and GIM2D
model parameters.  Second, in cases where PSF subtraction was
unsatisfactory, we fit a two--dimension smooth surface to the slight
ramping of the background (due to the quasar illumination) around a
$60 \times 60$ pixel region centered on the galaxy.

In two cases (Q0827+243 and Q1148+387), the post-- and pre-- PSF
subtracted asymmetries were consistent within the errors obtained via
the four realizations of the background. In four cases (Q0002+051,
Q0235+164, Q1241+176, and Q1246-057), the asymmetries differed by more
than their respective background induced errors (which are inherently
underestimates in that Poisson noise in the galaxy counts are not
included).  However, this was not a systematic effect. In two of those
cases (Q0002+051 and Q0235+164) the $A$ values increased and in two
cases (Q1241+176 and Q1246-057) the $A$ values decreased.

\begin{deluxetable*}{llcccccc}
\tabletypesize{\scriptsize}
\tablecaption{Galaxy Asymmetries\label{tab:abs}}
\tablecolumns{8}
\tablewidth{0pt}
\tablehead{
\colhead{QSO Field}&
\colhead{$z_{abs}$} &
\colhead{$A$} &
\colhead{$R_{A}^{1\sigma, 2r_{h}}$} &
\colhead{$R_{A}^{3\sigma, 2r_{h}}$} &
\colhead{$ A_{z}^{2\sigma, 2r_{h}}$} &
\colhead{$ A_{z}^{3\sigma, 2r_{h}}$} &
\colhead{$D_z$\tablenotemark{a}} 
}
\startdata
Q$0002+051 $ & $ 0.298059 $ &$  0.096 \pm 0.0008 $ & $ 0.050  $  &  $ 0.040  $ & $ 0.040 $ & $  0.020 $ & $ 0.051 $ \\
Q$0002+051 $ & $ 0.591365 $ &$  0.071 \pm 0.0003 $ & $ 0.050  $  &  $ 0.010  $ & $ 0.020 $ & $  0.010 $ & $ 0.090 $ \\
Q$0002+051 $ & $ 0.851407 $ &$  0.093 \pm 0.011\phantom{0}  $ & $ 0.090  $  &  $ 0.040  $ & $ 0.010 $ & $  0.000 $ & $ 0.220 $ \\
Q$0109+200 $ & $ 0.5346   $ &$  0.041 \pm 0.0053 $ & $ 0.042  $  &  $ 0.016  $ & $ 0.013 $ & $  0.000 $ & $ 0.236 $ \\
Q$0117+213 $ & $ 0.729075 $ &$  0.079 \pm 0.0018 $ & $ 0.060  $  &  $ 0.050  $ & $ 0.010 $ & $  0.000 $ & $ 0.130 $ \\
Q$0150-202 $ & $ 0.3887   $ &$  0.095 \pm 0.0012 $ & $ 0.058  $  &  $ 0.052  $ & $ 0.069 $ & $  0.028 $ & $ 0.075 $ \\
Q$0229+131 $ & $ 0.417337 $ &$  0.089 \pm 0.0008 $ & $ 0.080  $  &  $ 0.070  $ & $ 0.050 $ & $  0.030 $ & $ 0.060 $ \\
Q$0235+164 $ & $ 0.5240   $ &$  0.210 \pm 0.026  $ & $ 0.085  $  &  $ 0.038  $ & $ 0.005 $ & $  0.002 $ & $ \cdots $ \\
Q$0450-132 $ & $ 0.493937 $ &$  0.125 \pm 0.0010 $ & $ 0.110  $  &  $ 0.110  $ & $ 0.050 $ & $  0.040 $ & $ 0.050 $ \\
Q$0450-132 $ & $ 0.493937 $ &$  0.202 \pm 0.0005 $ & $ 0.190  $  &  $ 0.190  $ & $ 0.100 $ & $  0.080 $ & $ 0.050 $ \\
Q$0827+243 $ & $ 0.524968 $ &$  0.093 \pm 0.0060 $ & $ 0.100  $  &  $ 0.110  $ & $ 0.050 $ & $  0.020 $ & $ 0.100 $ \\
Q$0836+113 $ & $ 0.7874   $ &$  0.168 \pm 0.0017 $ & $ 0.220  $  &  $ 0.210  $ & $ 0.170 $ & $  0.130 $ & $ 0.210 $ \\
Q$1019+309 $ & $ 0.3461   $ &$  0.079 \pm 0.0011 $ & $ 0.070  $  &  $ 0.070  $ & $ 0.030 $ & $  0.010 $ & $ 0.020 $ \\
Q$1038+064 $ & $ 0.441453 $ &$  0.085 \pm 0.0006 $ & $ 0.080  $  &  $ 0.080  $ & $ 0.030 $ & $  0.010 $ & $ 0.030 $ \\
Q$1127-145 $ & $ 0.312710 $ &$  0.111 \pm 0.0013 $ & $ 0.095  $  &  $ 0.091  $ & $ 0.016 $ & $  0.007 $ & $ 0.072 $ \\
Q$1127-145 $ & $ 0.312710 $ &$  0.138 \pm 0.0004 $ & $ 0.073  $  &  $ 0.082  $ & $ 0.127 $ & $  0.054 $ & $ 0.209 $ \\
Q$1148+387 $ & $ 0.553362 $ &$  0.055 \pm 0.0025 $ & $ 0.090  $  &  $ 0.090  $ & $ 0.080 $ & $  0.040 $ & $ 0.350 $ \\
Q$1209+107 $ & $ 0.392925 $ &$  0.155 \pm 0.0022 $ & $ 0.107  $  &  $ 0.080  $ & $ 0.080 $ & $  0.061 $ & $ 0.063 $ \\
Q$1222+228 $ & $ 0.550196 $ &$  0.079 \pm 0.0018 $ & $ 0.100  $  &  $ 0.090  $ & $ 0.050 $ & $  0.020 $ & $ 0.100 $ \\
Q$1241+176 $ & $ 0.550483 $ &$  0.061 \pm 0.012\phantom{0}  $ & $ 0.050  $  &  $ 0.020  $ & $ 0.060 $ & $  0.030 $ & $ 0.100 $ \\
Q$1246-057 $ & $ 0.639908 $ &$  0.122 \pm 0.017\phantom{0}  $ & $ 0.160  $  &  $ 0.170  $ & $ 0.220 $ & $  0.070 $ & $ 0.860 $ \\
Q$1317+277 $ & $ 0.660053 $ &$  0.135 \pm 0.0010 $ & $ 0.080  $  &  $ 0.080  $ & $ 0.080 $ & $  0.050 $ & $ 0.020 $ \\
Q$1332+552 $ & $ 0.3740   $ &$  0.105 \pm 0.0006 $ & $ 0.050  $  &  $ 0.040  $ & $ 0.020 $ & $  0.020 $ & $ 0.050 $ \\
Q$1511+103 $ & $ 0.4369   $ &$  0.053 \pm 0.0017 $ & $ 0.070  $  &  $ 0.070  $ & $ 0.040 $ & $  0.030 $ & $ 0.040 $ \\
Q$1622+235 $ & $ 0.317596 $ &$  0.144 \pm 0.0053 $ & $ 0.050  $  &  $ 0.030  $ & $ 0.010 $ & $  0.000 $ & $ 0.010 $ \\
Q$1622+235 $ & $ 0.368113 $ &$  0.078 \pm 0.0050 $ & $ 0.080  $  &  $ 0.080  $ & $ 0.030 $ & $  0.030 $ & $ 0.010 $ \\
Q$1622+235 $ & $ 0.471930 $ &$  0.143 \pm 0.0052 $ & $ 0.060  $  &  $ 0.080  $ & $ 0.040 $ & $  0.040 $ & $ 0.010 $ \\
Q$1622+235 $ & $ 0.656103 $ &$  0.037 \pm 0.0059 $ & $ 0.000  $  &  $ 0.020  $ & $ 0.020 $ & $  0.010 $ & $ 0.030 $ \\
Q$1622+235 $ & $ 0.702903 $ &$  0.142 \pm 0.0052 $ & $ 0.116  $  &  $ 0.105  $ & $ 0.104 $ & $  0.076 $ & $ 0.019 $ \\
Q$1622+235 $ & $ 0.797079 $ &$  0.091 \pm 0.0055 $ & $ 0.080  $  &  $ 0.060  $ & $ 0.070 $ & $  0.040 $ & $ 0.050 $ \\
Q$1622+235 $ & $ 0.891280 $ &$  0.080 \pm 0.019\phantom{0}  $ & $ 0.110  $  &  $ 0.130  $ & $ 0.060 $ & $  0.020 $ & $ 0.220 $ \\
Q$1623+268 $ & $ 0.8885   $ &$  0.071 \pm 0.0050 $ & $ 0.100  $  &  $ 0.060  $ & $ 0.050 $ & $  0.030 $ & $ 0.140 $ \\
Q$1623+268 $ & $ 0.8875   $ &$  0.050 \pm 0.0058 $ & $ 0.100  $  &  $ 0.090  $ & $ 0.560 $ & $  0.240 $ & $ \cdots $ \\
Q$2128-123 $ & $ 0.429820 $ &$  0.036 \pm 0.0029 $ & $ 0.036  $  &  $ 0.042  $ & $ 0.021 $ & $  0.009 $ & $ 0.250 $ \\
Q$2206-199 $ & $ 0.751923 $ &$  0.202 \pm 0.0002 $ & $ 0.135  $  &  $ 0.035  $ & $ 0.061 $ & $  0.062 $ & $ 0.036 $ \\
Q$2206-199 $ & $ 0.948361 $ &$  0.089 \pm 0.0017 $ & $ 0.064  $  &  $ 0.041  $ & $ 0.009 $ & $  0.002 $ & $ 0.043 $ \\
Q$2206-199 $ & $ 1.017040 $ &$  0.306 \pm 0.0024 $ & $ 0.460  $  &  $ 0.530  $ & $ 0.430 $ & $  0.310 $ & $ 0.670 $ \\
\enddata
\tablenotetext{a}{The $D_z$ computations for the galaxies Q$0235+164$
at $z=0.5240$ and Q$1623+268$ at $z=0.8875$ failed.}
\end{deluxetable*}

Due to the inherent uncertainties in PSF subtraction of saturated
quasars, we adopt the $A$ values from the pre--PSF subtracted images
and use the deviations between the pre-- and post-- PSF subtracted images
as a measure of the uncertainty in the $A$ values. The adopted
uncertainties for these seven galaxy--absorber pairs are roughly a
factor of five larger than their background only induced errors.

\section{Analysis and Results}
\label{sec:results}

In Table~\ref{tab:abs}, we present the quantified asymmetries for each
absorption selected galaxy in our sample.  In columns (1) and (2), we
list the quasar field and the absorption redshift.  In column (3), we
list the Abraham $A$ and uncertainty.  For purposes of later
discussion, we list representative values of $R_A$ and $A_z$, for
apertures $2~r_h$ and for $3~\sigma$ and $2~\sigma$ significance
thresholds, respectively, in columns (4)--(5) and (6)--(7).  In column
(8), we list $D_z$.

In order to explore possible correlations between galaxy asymmetries
and the strength of the absorbing gas, we performed non--parametric
Spearman and Kendall rank correlation tests between the various
asymmetry measures, $A$, $R_A$, $A_z$, and $D_z$, and the {\MgII}
$\lambda 2796$ equivalent width.  We also examined various scalings of
the asymmetries, most notably normalization by the impact parameter,
$D$.

For all tests, the null--hypothesis of no correlation could not be
ruled out to a high level of significance (i.e., greater than
$3~\sigma$), except for the Abraham $A$, for which a 3.1~$\sigma$
correlation was found between $A/D$ and $W_r(2796)$.  Recall that the
Abraham $A$ is model independent and is the only asymmetry based upon
an elliptical aperture, whereas the $R_A$, $A_z$, $D_z$ are derived
from the circular model dependent apertures at various $r_h$.  Thus, $A$
represents a full galaxy, model independent asymmetry with a
natural isophotal aperture.

At face value, the correlation between $A/D$ and $W_r(2796)$ for the
full sample (see the inset of Figure~\ref{fig:properties}) suggests
that, once the impact parameter of absorption is normalized out, there
may be a connection between perturbations in a galaxy's morphology and
the quantity of {\MgII} absorbing gas probed in its halo (which is a
complex distribution of column density and velocity spread).  Note
that for the largest equivalent widths, the scatter in $A/D$ increases
substantially.  Based upon this impression, we might speculate that
there is a physically based transition in the behavior of $W_r(2796)$
for large equivalent widths, i.e., a transition from a ``clouds in
halos'' scenario to a ``galactic winds'' scenario [see, for example,
\citet{bond01b}].  Since absorption profiles become progressively more
saturated with increasing equivalent width, it is more conservative to
simply suggest that this increase in scatter could arise due to a loss
of information on the quantity of gas probed by the quasar.

\subsection{Equivalent Widths and Saturation}

The equivalent width is a single valued measure of a complex velocity
distribution of gas column density.  For ``weak'' and ``classic''
systems, distinct individual clouds are usually detected with low to
moderate column densities, and the number of clouds strongly
correlates with equivalent width \citep{cv01,cvc03}.  However, for
much ``stronger'' systems, the correlation breaks down for saturated
profiles, indicating that the equivalent width can no longer be
employed to accurately decipher the combination of column density,
number of clouds, and their velocity distribution.

\begin{figure}[b]
\figurenum{2}
\plotone{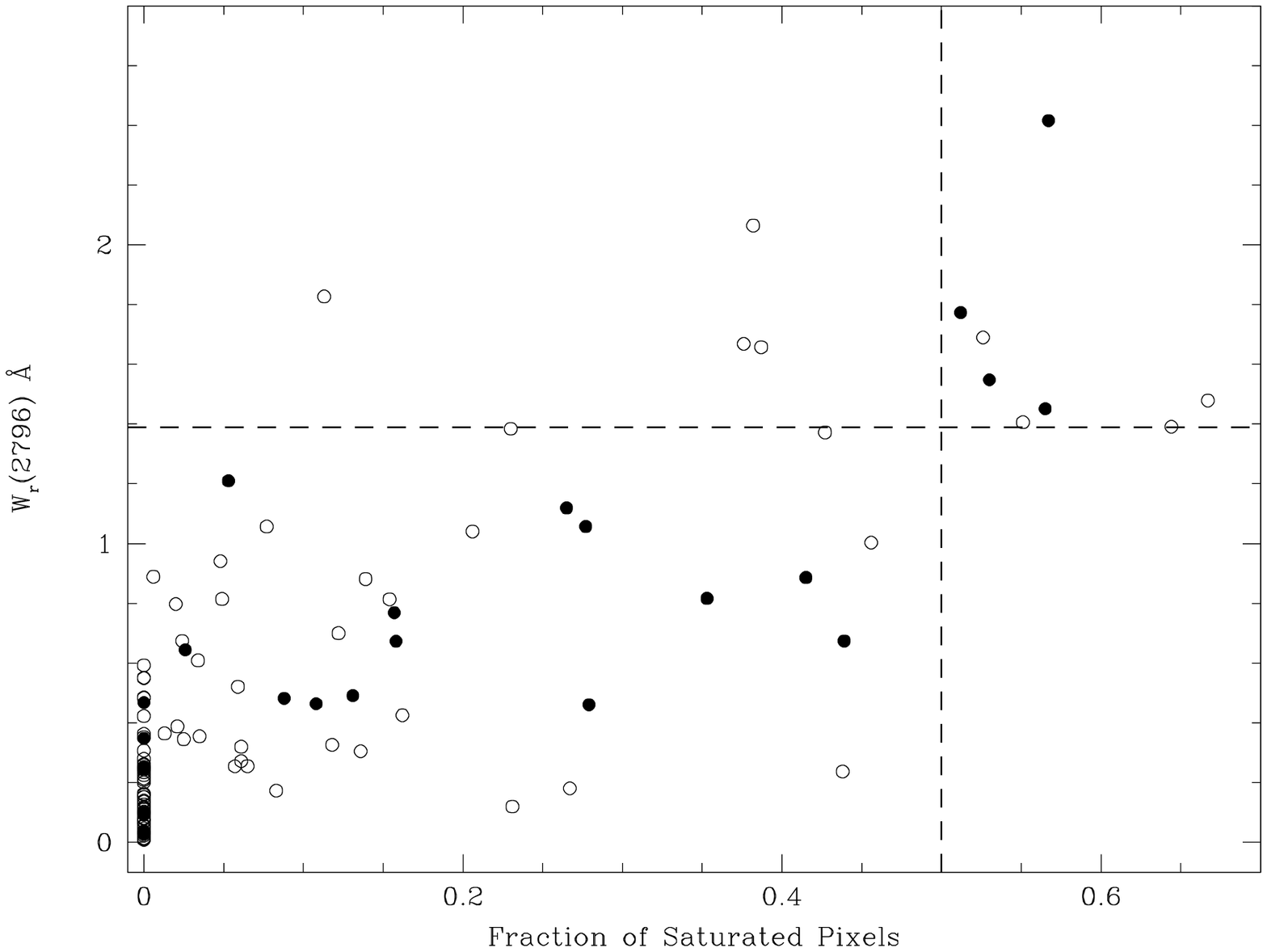}
\caption{The {\MgII} $\lambda 2796$ equivalent width as a function of
the ratio of the number of saturated pixels to total pixels, $f_{sat}$,
for profiles observed with HIRES and UVES spectra. Both the dashed
lines were obtained by maximizing the number of points in the lower
left and upper right quadrants (Q1 and Q3) and minimizing the number
of points in lower right and upper left quadrants (Q2 and Q4).  We
define a profile as highly saturated for $f_{sat} \geq 0.5$.  The
horizontal dashed line provides a natural and objective equivalent
width transition of $W_r(2796) = 1.39$~{\AA} for defining our subsample.}
\label{fig:sat}
\end{figure}

Using an approach similar to the saturation $D$--index employed by
\citet{ellison06} to efficiently pre--selected damped {\Lya} absorbers
(DLAs), we computed the fraction saturated pixels, $f_{sat}$, across
each HIRES or UVES {\MgII} $\lambda 2796$ absorption profile.  The
full pixel range of a profile is used, as defined using the methods of
\citet{cv01}.  We use the criterion that a pixel is saturated when the
counts are less than 1.5~$\left< \sigma \right>$, where $\left< \sigma
\right>$ is the average of the uncertainty spectrum computed over a
window $\pm 600$~{\kms} from the optical depth mean (which defines the
velocity zero point).  In Figure~\ref{fig:sat}, we present $W_r(2796)$
versus the fraction of saturated pixels, $f_{sat}$.  The solid points
represent the galaxies in our sample for which we have HIRES and UVES
spectra of the {\MgII} $\lambda 2796$ absorption profiles.  For these
data, there is some suggestion that the diagram visually breaks into
four quadrants divided by $W_r(2796) \simeq 1.3$~{\AA} and $f_{sat}
\simeq 0.5$.  These quadrants tend to be either fully populated or
entirely void of data points.  

In order to ascertain whether this trend is universal to HIRES and
UVES {\MgII} absorption profiles, we populated the figure with an
additional 97 {\MgII} systems from our database of HIRES and UVES
spectra.  These systems were selected to cover the same redshift and
equivalent width ranges of our sample and will be presented elsewhere
as part of a larger survey \citep{evans07}.  We plot these
supplementary data as open circles in Figure~\ref{fig:sat}.  As can be
seen, the general trend remains, especially the void of data in the
quadrant defined by $W_r(2796) \leq 1.4$~{\AA} and $f_{sat} \geq 0.5$.
The behavior of {\MgII} absorption is such that there are no profiles
with a saturation fraction of $f_{sat} \geq 0.5$ and equivalent width
less than $W_r(2796) \simeq 1.4$~{\AA}.  For equivalent widths above
this value, there is a small population of profiles with $f_{sat} <
0.5$, but none of these are present in our sample.

We partitioned the $W_r(2796)$--$f_{sat}$ plane into four quadrants in
order to objectively classify the different types of absorption
profiles with the goal of defining a subsample unaffected by strong
saturation.  We refer to these as Q1--Q4, starting at the origin and
increasing counterclockwise on the plane (Q2 is the empty quadrant).
The final quadrants were determined by maximizing the number of data
points in Q1 and Q3 and minimizing the number in Q2 and Q4.  This
method resulted in a strict equivalent width transition of
$W_r(2796)=1.39$~{\AA} and a less stringent transition of saturation
in the range $0.46 \leq f_{sat} \leq 0.51$.  The dashed lines on
Figure~\ref{fig:sat} define the resulting quadrants, where we have
adopted $f_{sat} \geq 0.5$ to designate highly saturated profiles.  A
Kolomorov--Smirnov (K--S) test yielded a probability of $P_{_{\rm KS}}
< 10^{-8}$ ($5.2~\sigma$) that systems with $W_r(2796)<1.39$~{\AA} and
$W_r(2796)\geq 1.39$~{\AA} are drawn from the same distribution of
$f_{sat}$.

There are a total of nine galaxy--absorber pairs in our sample that
fall into our ``highly saturated'' classification.  For four
(associated with five galaxies), we have HIRES and UVES spectra and we
have directly computed their $f_{sat}$.  However, there are four in
the range $2.3 \leq W_r(2796) \leq 2.90$~{\AA} for which we do not
have HIRES or UVES quasar spectra (Q$0109+200$, Q$0235+164$,
Q$0836+113$, and Q$1332+552$), and cannot directly compute their
$f_{sat}$.  Fortunately, the {\MgII} $\lambda 2796$ profiles of the
absorbers in the Q$0235+164$ and Q$0836+113$ fields are available for
inspection in \citet{lanzetta92} and \citet{turnshek89}, respectively.
Their spectra are of comparable resolution to HIRES and UVES.  These
two absorbers appear to clearly have $f_{sat} > 0.5$, indicating that
they are also highly saturated.  Based upon our investigation, it is
reasonable to assume that the very large $W_r(2796)$ absorbers in the
Q$0109+200$ and Q$1332+552$ fields also have highly saturated
profiles.

\subsection{Examining a Homogeneous Subsample}

Applying the above motivated 50\% saturation limit, which corrersponds
to $W_r(2796) < 1.39$~{\AA}, we obtain a subsample of galaxy--absorber
pairs with a higher degree of homogeneity that can be investigated in
more detail for a deeper understanding of the nature of
$A/D$--$W_r(2796)$ correlation.  The subsample comprises 28
absorber--galaxy pairs from which the highly saturated profiles have
been excluded.  For these highly saturated profiles, the single valued
equivalent width is a less reliable measure of the quantity and
complex distribution of absorbing gas.

As with the full sample, we performed non--parametric Spearman and
Kendall correlation tests with the computed asymmetries for the
subsample of 28.  In addition to the equivalent widths, we included
high resolution kinematic properties of the absorption, i.e., the
velocity spread, velocity asymmetry, number of clouds (based upon
Voigt profile decomposition), and the total {\MgII} column density
[see \citet{cv01} and \citet{cvc03} for definitions of these
quantities].  Though these kinematic quantities correlate with
$W_r(2796)$ to varying degrees (only the number of clouds is
statistically significant), no single kinematic quantity provides a
unique description of the magnitude of absorption as measured by the
equivalent width \citep{cv01}.  Interestingly, none of the kinematic
absorption properties were found to correlate with the various
asymmetry measures, nor with the asymmetries normalized by impact
parameter.  However, the significance of the $A/D$--$W_r(2796)$
correlation remains highly significant with a confidence level (CL) of
99.9\% (a $3.33~\sigma$ result that the null--hypothesis of no
correlation is ruled out).

We explored the statistical behavior of the correlations between
$W_r(2796)$ and all the asymmetry measures computed using GIM2D.
Recall that the Abraham $A$ is computed over an elliptical aperture
defined by $2~\sigma$ above background, whereas $R_A$ and $A_z$ are
computed for circular apertures at 10 values of $r_h$ (model
dependent) and for three different levels of significance.  A selected
subset of our exploration is presented in Table~\ref{tab:kstest2}.  In
column (1) are the tested properties. In columns (2)--(4) are the
Spearman correlation coefficient, $r_s$, the probability, $P_S$, that
the tested data are consistent with the null hypothesis, and the number
of standard deviations, $N_{\sigma}$ assuming the ranks are drawn from
a normal distribution. In columns (5)--(7) are the Kendall $\tau$, the
probability, $P_K$, and $N_{\sigma}$.  For simplicity, we will
hereafter discuss the Kendall statistics.

In Table~\ref{tab:kstest2}, we first report that there is no
significant correlation (or anti--correlation) between impact
parameter, $D$, and $W_r(2796)$ for our subsample.  The equivalent
width does trend toward smaller values as $D$ increases, but the CL is
92.5\% ($1.8~\sigma$), which does not classify as a statistically
significant departure from the null--hypothesis.  However, in our
exploration, we consider the possible affects of the trend between $D$
and $W_r(2796)$.

\begin{figure}[h]
\figurenum{3}
\plotone{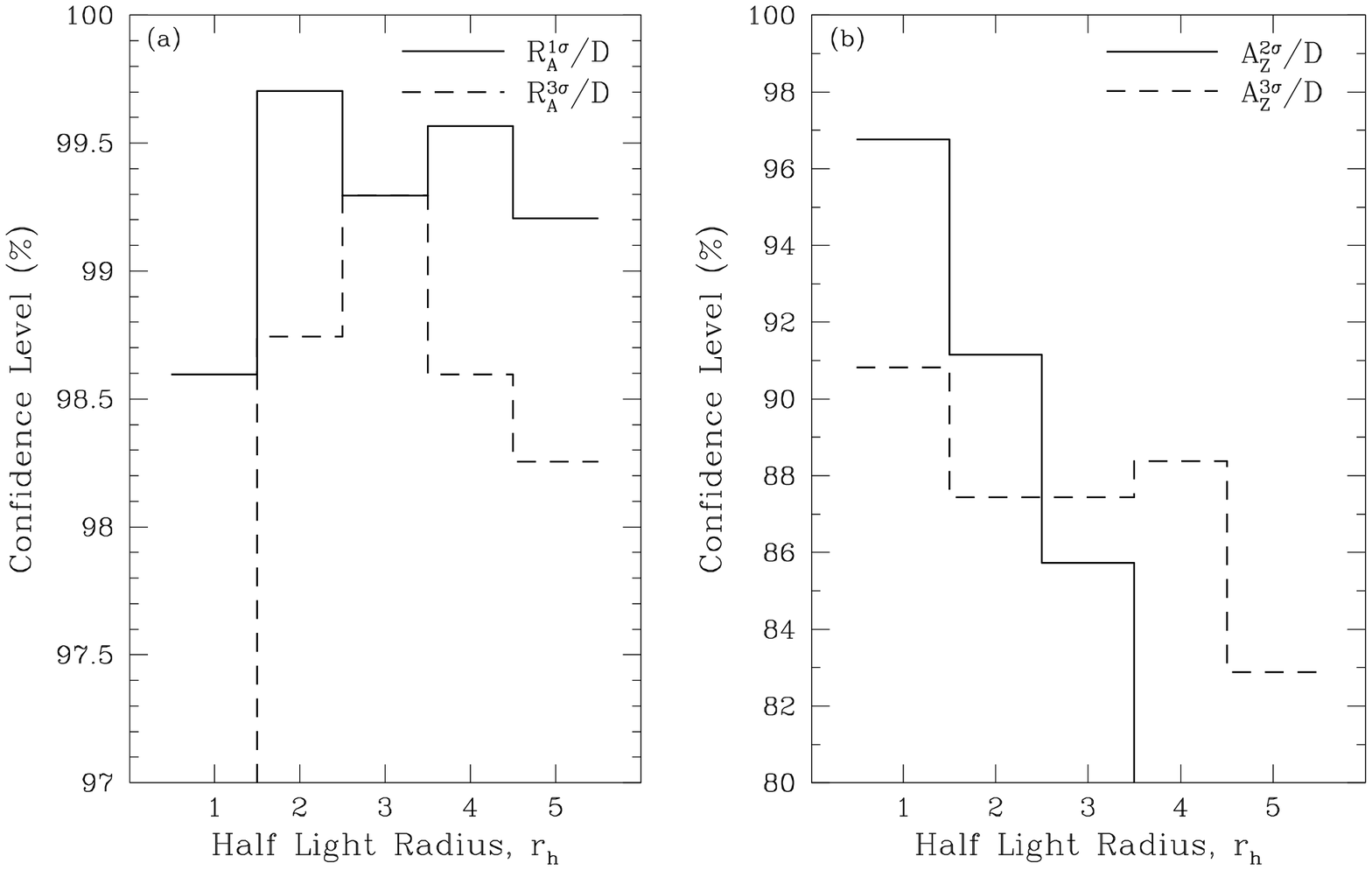}
\caption{($a$) Confidence level (CL) for the correlation between
$W_r(2796)$ and $R_A/D$ over the range $1 \leq r_{h} \leq 5$ for
1~$\sigma$ (solid line) and 3~$\sigma$ (dashed line) thresholds above
background. $R_A$ is computed from the model residual images, which
are less robust near the galaxy core ($r_{h}\le 1$). --- ($b$) CL for
the correlation between $W_r(2796)$ and $A_z/D$ over the range $1 \leq
r_{h} \leq 5$ for the 2~$\sigma$ (solid line) and 3~$\sigma$ (dashed
line) cases. $A_z$ is computed directly from the data.  Both $R_A$ and
$A_z$ employ the circular $r_h$ apertures of the models.  Note the
different CL scales for panels $a$ and $b$.}
\label{fig:compare}
\end{figure}

We first focus on the asymmetry $R_A$ for both $R_{A}^{3\sigma,
2r_{h}}$ and $R_{A}^{3\sigma, 3r_{h}}$ (3~$\sigma$ above background
for 2 and $3~r_{h}$, respectively).  As presented in
Table~\ref{tab:kstest2}, there is a total lack of a correlation for
each of $W_r(2796)$ versus $R_{A}^{3\sigma, 2r_{h}}$ and $W_r(2796)$
versus $R_{A}^{3\sigma, 3r_{h}}$ with $\tau_{_{\rm K}} = 0.07$ and
$0.06$, respectively.  These statistics translate to confidence levels
(CL) of 40.2\% and 34.5\%.  When the $R_A$ are normalized by $D$,
moderately significant correlations arise at the 98.7\% (2.5~$\sigma$)
and 99.3\% (2.7~$\sigma$) CL, respectively.  To investigate these
results as a function of the number of $r_h$, we present the behavior
of the CL for both $R_{A}^{1\sigma}/D$ and $R_{A}^{3\sigma}/D$ as a
function of the number of half--light radii out to 5~$r_{h}$ in
Figure~\ref{fig:compare}$a$.  For both sensitivity thresholds, the CL
scatter around $\sim 99$\%, which are moderately significant
correlations.  The CL for the $R_A$ substantially decrease for
1~$r_{h}$. This is not unexpected since the $R_A$ are computed using
the model residual images; the GIM2D models have difficulties matching
the surface brightness profiles in the galaxy cores, resulting in
artifacts in the residual images for $r_{h} \leq 1$. These model
dependent artifacts tend to randomize the $R_A$ in the cores, which
washes out any signal in the correlation tests.

We next list a selected correlation test for the asymmetry for
$A_{z}^{2\sigma, 1r_{h}}$ (pixels 2~$\sigma$ greater than their
symmetric counterparts within 1~$r_{h}$) and the correlation test for
$D_z$ in Table~\ref{tab:kstest2}.  As with $R_A$, when the $A_z$ is
normalized by $D$, the Kendall probability increases from a
statistically insignificant CL of 79.6\% (1.3~$\sigma$) to a
marginally significant 96.8\% (2.1~$\sigma$).  In
Figure~\ref{fig:compare}$b$, we show the CL behavior of
$A_{z}^{2\sigma}/D$ and $A_{z}^{3\sigma}/D$ as a function of the
number of half--light radii out to 5~$r_{h}$. The $A_z$ are computed
directly on the data; there are no artifacts at $r_h \leq 1$
introduced by the GIM2D models.  The CL of the $A_{z}^{2\sigma}/D$ and
$A_{z}^{3\sigma}/D$ tests substantially decrease for increasing
$r_{h}$.  The correlation test for $D_{z}$ and $W_{r}(2796)$ reveals
no indication of even a slight trend (yielding $\tau_{_{\rm K}} \simeq
0.02$) even when $D_z$ is normalized by $D$ (yielding $\tau_{_{\rm K}}
\simeq 0.08$).  Recall that $D_{z}$ is sensitive to the presence of
tidal arms for $r_h > 1$.

\begin{deluxetable}{lccccccc}
\tabletypesize{\footnotesize}
\tablecolumns{8} 
\tablewidth{0pt} 
\tablecaption{Results of K--S Tests With $W_{r}(2796)<1.39$~{\AA} \label{tab:kstest2}}

\tablehead{ 
\colhead{ }  &
\multicolumn{3}{c}{Spearman} & &
\multicolumn{3}{c}{Kendall} \\
\cline{2-4} \cline{6-8} 
\colhead{Tests} & 
\colhead{$r_{_{\rm S}}$} & 
\colhead{$P_{_{\rm S}}$}  & 
\colhead{$N_{_{\sigma}}$} & &
\colhead{$\tau_{_{\rm K}}$} & 
\colhead{$P_{_{\rm K}}$} & 
\colhead{$N_{_{\sigma}}$} 
}
\startdata 
$W_{r}$ vs. $D$                     & $-$0.31 & 0.1137 & 1.58 & & $-$0.24 & 0.0746 & 1.78\\
\cutinhead{$R_A$ Method \citep{schade95}}
$W_{r}$ vs. $R_{A}^{3\sigma, 2r_{h}}$   &  0.06 & 0.6798 & 0.41 & &  0.07 & 0.5977 & 0.53\\
$W_{r}$ vs. $R_{A}^{3\sigma, 3r_{h}}$   &  0.09 & 0.5713 & 0.57 & &  0.06 & 0.6552 & 0.45\\
                                        &       &        &      & &       &        &     \\[-2.0ex]
$W_{r}$ vs. $R_{A}^{3\sigma, 2r_{h}}/D$ &  0.45 & 0.0181 & 2.36 & &  0.33 & 0.0126 & 2.50\\
$W_{r}$ vs. $R_{A}^{3\sigma, 3r_{h}}/D$ &  0.50 & 0.0095 & 2.59 & &  0.36 & 0.0071 & 2.69\\
\cutinhead{$A_z$ and $D_z$ Methods \citep{simard02}}
$W_{r}$ vs. $ A_{z}^{2\sigma, 1r_{h}}$  &  0.23 & 0.2082 & 1.26 & &  0.17 & 0.2041 & 1.27\\
$W_{r}$ vs. $ D_z$                      &  0.03 & 0.8319 & 0.21 & &  0.02 & 0.8885 & 0.14\\
                                        &       &        &      & &       &        &     \\[-2.0ex]
$W_{r}$ vs. $A_{z}^{2\sigma, 1r_{h}}/D$ & 0.43  & 0.0243 & 2.25 & &  0.29 & 0.0324 & 2.14\\ 
$W_{r}$ vs. $ D_z/D$                    & 0.13  & 0.5029 & 0.67 & &  0.08 & 0.5523 & 0.59\\
\cutinhead{  $A$ Method \citep{abraham94}}
$W_{r}$ vs. $A$        &  0.40 & 0.0359 & 2.10 & &  0.28 & 0.0394 & 2.06\\ 
                                     &       &       &      & &       &       &     \\[-2.0ex]
$D$\phantom{$_r$} vs. $A$            &  0.25 & 0.1889 & 1.31 & &  0.20 & 0.1437 & 1.46\\  
                                     &       &       &      & &       &       &     \\[-2.0ex]
$W_{r}$ vs. $A/D$      &  0.64 & 0.0008 & 3.36 & &  0.45 & 0.0009 & 3.33\\
\enddata 
\end{deluxetable}

The different $r_h$ behavior of the CL for the $R_A$ and $A_z$
correlation tests can be explained by considering the effect of
Poisson noise. $R_A$ is sensitive to significant fluctuations about a
smooth, correlated, axisymmetric model surface. As the $r_h$ apertures
increase, the Poisson fluctuations in the pixel values become
relatively more important, but at a rate much slower than for that of
$A_z$. This is because $A_z$ is based upon differences between pixel
values and the Poisson fluctuations at larger $r_h$ apertures can wash
out the significance of these differences.

\begin{figure}
\figurenum{4}
\plotone{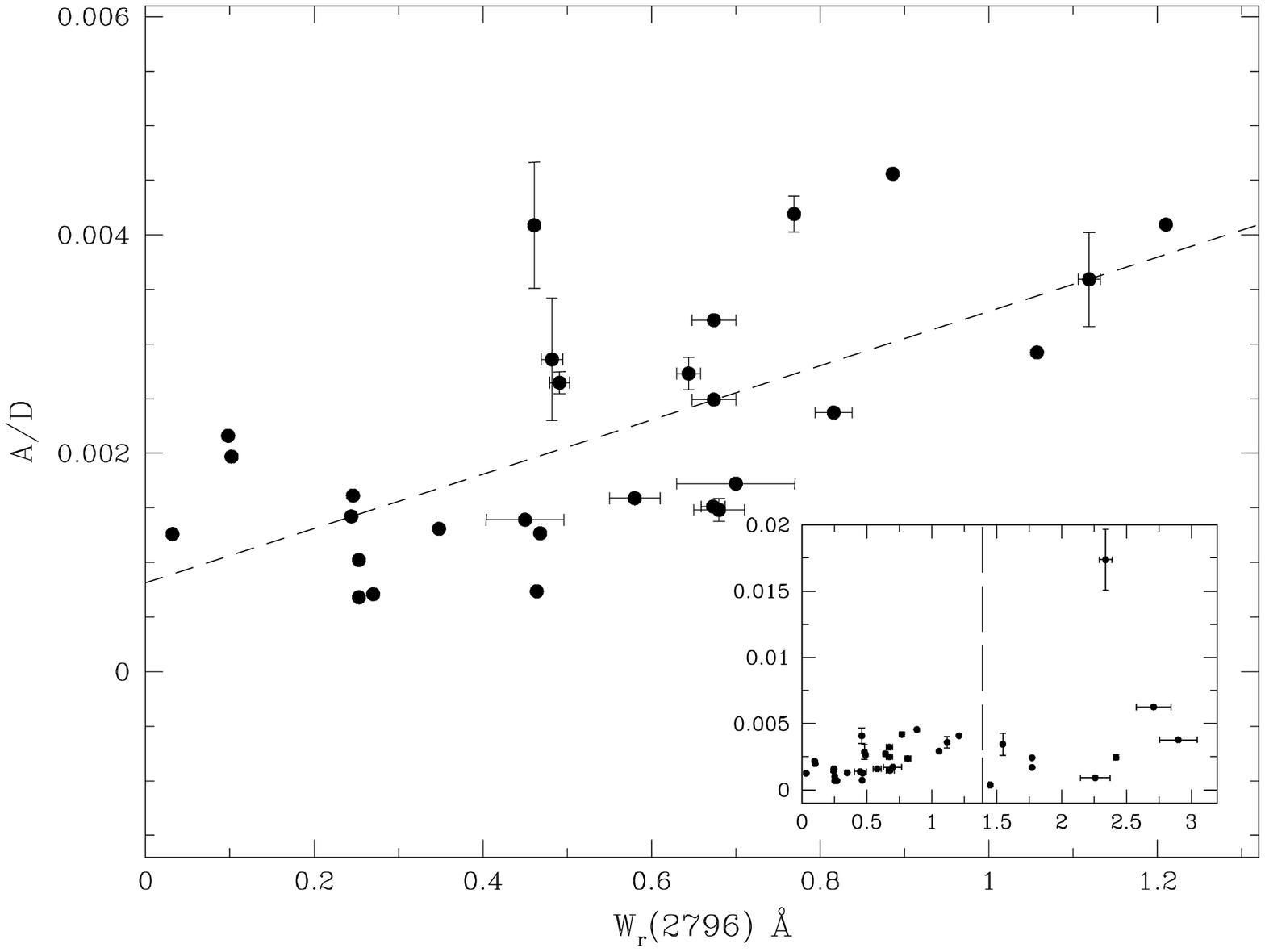}
\caption{Galaxy morphological asymmetry $A$ normalized to impact
  parameter, $D$, versus $W_r(2796)$ for the subsample with $f_{sat} <
  0.5$. The error bars are $1~\sigma$; in most cases, the errors are
  smaller than the data points. The correlation has a 99.92\%
  ($3.3~\sigma$) confidence level. The dashed line is a maximum
  likelihood fit, $A/D = 2.483\times 10^{-3} W_r(2796) + 8.15 \times
  10^{-4}$. --- (inset) The same plot for the full sample,
  illustrating the increased scatter for larger equivalent widths. The
  dashed line is the cutoff for the subsample.}
\label{fig:properties}
\end{figure}

The correlation test for the Abraham $A$ with $W_r(2796)$ is also
listed in Table~\ref{tab:kstest2}.  There is a moderately significant
correlation between $W_r(2796)$ and $A$, with CL of 96\%
(2.1~$\sigma$). We find only a weak trend between $A$ and $D$
(1.5~$\sigma$). However, when the $A$ are normalized by $D$, a
significant correlation emerges with a CL of 99.9\% (3.3~$\sigma$).
Since the $A$ are computed within a single isophotal area, we cannot
examine the behavior of $A$ with $r_h$.  In
Figure~\ref{fig:properties}, we present $A/D$ versus $W_r(2796)$, with
the dashed line showing a maximum likelihood fit.

It could be argued that the correlation between $A/D$ and $W_r(2796)$
might be induced by the weak trend (1.8~$\sigma$) between $W_r(2796)$
and $D$. To investigate this, we selected correlation tests of other
GIM2D model galaxy parameters versus $W_r(2796)$ for which the CL from
the correlation tests were similar to the tests of the asymmetries
presented in Table~\ref{tab:kstest2}. We then divided these model
galaxy parameters by $D$ and reran the correlation tests.  None of the
scaling tests yielded an ``induced'' correlation (the highest
significance was $\sim 2.0~\sigma$). In fact, in all cases where $2.0
< N_{\sigma} < 2.4$ prior to normalizing by $D$, these scaling tests
yielded reduced significance.

To further investigate the confidence level we found for the
$A/D$--$W_r(2796)$ correlation, we applied the bootstrap replacement
program {\sc resampling} \citep{howell04} to the subsample.  We
performed 10,000 bootstrapped samples, each of $n = 28$, with random
replacement from the original data to determine if the correlation is
due to a chance realization of the data. The bootstrap statistic
corroborates the Spearman--Kendall statistics, yielding a 99\%
confidence level.

\section{Discussion}
\label{sec:discussion}

In our exploration of the morphological properties of galaxies
selected by {\MgII} absorption, we have found a statistically
significant correlation ($3.3~\sigma$) between the {\MgII} $\lambda
2796$ equivalent width and galaxy asymmetry, $A$, normalized by the
impact parameter.  Background fluctuations in the calculation of $A$
introduce only a 0.1~$\sigma$ spread in the significance level of the
correlation.  Additional uncertainties in the $A$ values due to PSF
subtraction of the quasar in the seven small impact parameter
galaxy--absorber pairs (see \S~\ref{sec:psf}) introduced only a
$0.2~\sigma$ spread in the significance level.

It is of interest to address the individual correlation test results
for $A$ versus $D$ and for $A$ versus $W_r(2796)$ as presented in
Table~\ref{tab:kstest2}. First, $A$ versus $D$ shows a weak positive
trend such that larger $A$ are found at larger $D$. This suggests that
uncertainty due to PSF subtraction of the quasar is not inducing
elevated $A$ values for galaxies at small $D$. Second, $A$ versus
$W_r(2796)$ exhibits a moderately significant correlation
(2.1~$\sigma$) in that large $W_r(2796)$ are associated with galaxies
measured to have larger $A$. This hints that there may be a {\it
direct} connection between galaxy morphological asymmetry and the
quantity of gas in the galaxy halo. 

However, as described above, normalization of $A$ by $D$ accentuates
that trend, yielding a significant correlation.  Therefore, for
absorption selected galaxies, the $A/D$--$W_r(2796)$ correlation
suggests a connection between the magnitude of asymmetric
perturbations in a galaxy's morphology, normalized to the
galactocentric distance probed by the quasar, and the quantity of
absorbing gas residing at that distance in its halo.  In effect, this
might be interpreted such that the influence of galaxy asymmetry
diminishes with distance in a halo. The significance of the
correlation is substantiated by the bootstrap replacement method.  We
note that this result is independent of the GIM2D galaxy models, since
$A$ is computed directly from the counts in the galaxy.

\subsection{Comparison of Asymmetry Measures}

The different computational details of $A$, $R_A$, $A_z$, and $D_z$
render each with varying degrees of sensitivity to the details of the
underlying structures that give rise to the morphological asymmetries in
the galaxies.  Furthermore, the relative behavior between the
correlation tests provide insight for interpreting the correlation.

The CL behavior of $A_z$ with half--light radius aperture, $r_h$,
suggests that the tendency for a correlation of $A_z/D$ with
$W_r(2796)$ arises in the centrally concentrated region of the
galaxies, primarily within $1~r_h$ (see Figure~\ref{fig:compare}).  In
contrast, the CL behavior of $R_A$ is fairly flat as a function of the
$r_h$ circular aperture.  For larger $r_h$ apertures, the sensitivity
of $R_A$ to asymmetries (based upon {\it model residuals\/} and the
parameter $r_h$) may be higher than that of $A_z$ (based upon the {\it
data\/} and the parameter $r_h$).  However, since $R_A$ is an entirely
model dependent quantity, it is not highly reliable in the central
aperture ($r_h = 1$) due to difficulties with the model in the galaxy
cores (yielding poor subtraction of the galaxy model in the core and
thus resulting in a low CL).  The low CL of $D_z$, which is computed
exclusively outside the $r_h = 1$, suggests that the asymmetries are
not due to strong spiral arms or tidal arms to one side of the galaxy
(which is both consistent with the behavior of $A_z$ and visual
inspection of the galaxy images).

On the other hand, the Abraham $A$ is an entirely model independent
quantity.  It is computed directly on the data over a generalized
elliptical aperture defined by the region where galaxy counts are
$2~\sigma$ above background.  Thus, the $A/D$--$W_r(2796)$ correlation
is independent of any arbitrary assumption (such as the functional
form of the bulge and disk surface brightness profiles) applied in the
GIM2D modeling.  Since the apertures are operationally defined using
the signal--to--noise ratio of the galaxies, the asymmetries could in
principle be sensitive to the pixel counts of each galaxy. The pixel
counts for a given galaxy depend upon the product of exposure time and
surface brightness.  We calculated the signal--to--noise ratio from
the surface brightness of each galaxy over their full $1.5~\sigma$
aperture. A correlation test yielded no trend between $A$ and galaxy
signal--to--noise ratio. 

The combined behavior of the asymmetries and the correlation tests
with $r_h$ indicate that the asymmetries of the galaxies in our sample
are centrally concentrated.  The asymmetries are not the result of
major perturbations that would result in significant departures from a
de Vaucouleurs profile in the bulge components and an exponential
profile in the disk components.  As such, the morphologies of {\MgII}
absorption selected galaxies appear to be ``normal'' in their
classifications.  This suggests that extraordinary events, such as
major galaxy merging or a history of galaxy harassment, are not
characteristic of galaxies hosting extended gaseous halos.

\subsection{Asymmetries and Absorption Selection Galaxies}

Since the galaxies in our study are selected by association with
{\MgII} absorption, it is of interest to know if this selection
criteria results in a population of galaxies with higher levels of
morphological asymmetries (or perturbations) than the general
population of magnitude selected galaxies at intermediate redshifts.
The Medium Deep Survey \citep[MDS,][]{abraham96b}, which employs the
F814W filter, is the only survey to which we can undertake a fair
comparison since the MDS galaxies have also been modeled using GIM2D
and are drawn from both a similar magnitude range ($20<m_{F814W}<22$)
and redshift range as our sample. \citet{crampton95} argue that the
majority of galaxies in the MDS reside at $z<1.0$ with a peak in the
distribution at $z\simeq0.6$.  The range of magnitudes of our sample
is $19.4<m_{F702W}<23.8$ and covers the redshift interval $0.3<z <1$
with a peak in the distribution at $z=0.43$.

It is well known that the details of galaxy morphologies depend upon
the rest--frame bandwidth in which they are observed.  Almost all of
the galaxies in our sample were observed with the F702W filter; only
two have been observed in the F814W filter, whereas the MDS galaxies
are all observed in the F814W filter.  Thus, there is some concern
that a quantitative comparison of asymmetries between our sample
galaxies with those of the MDS may be compromised due to a
non--uniform rest--frame band in which the galaxies are observed.  We
can only address this concern statistically.  Since the peak of the
redshift distribution of MDS galaxies is $z\sim$~0.6, the majority of
the galaxies are observed in the rest--frame wavelength band
4530--5490~{\AA}.  For our sample, the redshift distribution peaks at
$z=0.43$, which translates to a rest--frame wavelength band of
4360--5340~{\AA} for the majority of our sample.  Therefore, there is
a significant rest--frame wavelength overlap between the two samples.

\begin{figure}
\figurenum{5}
\plotone{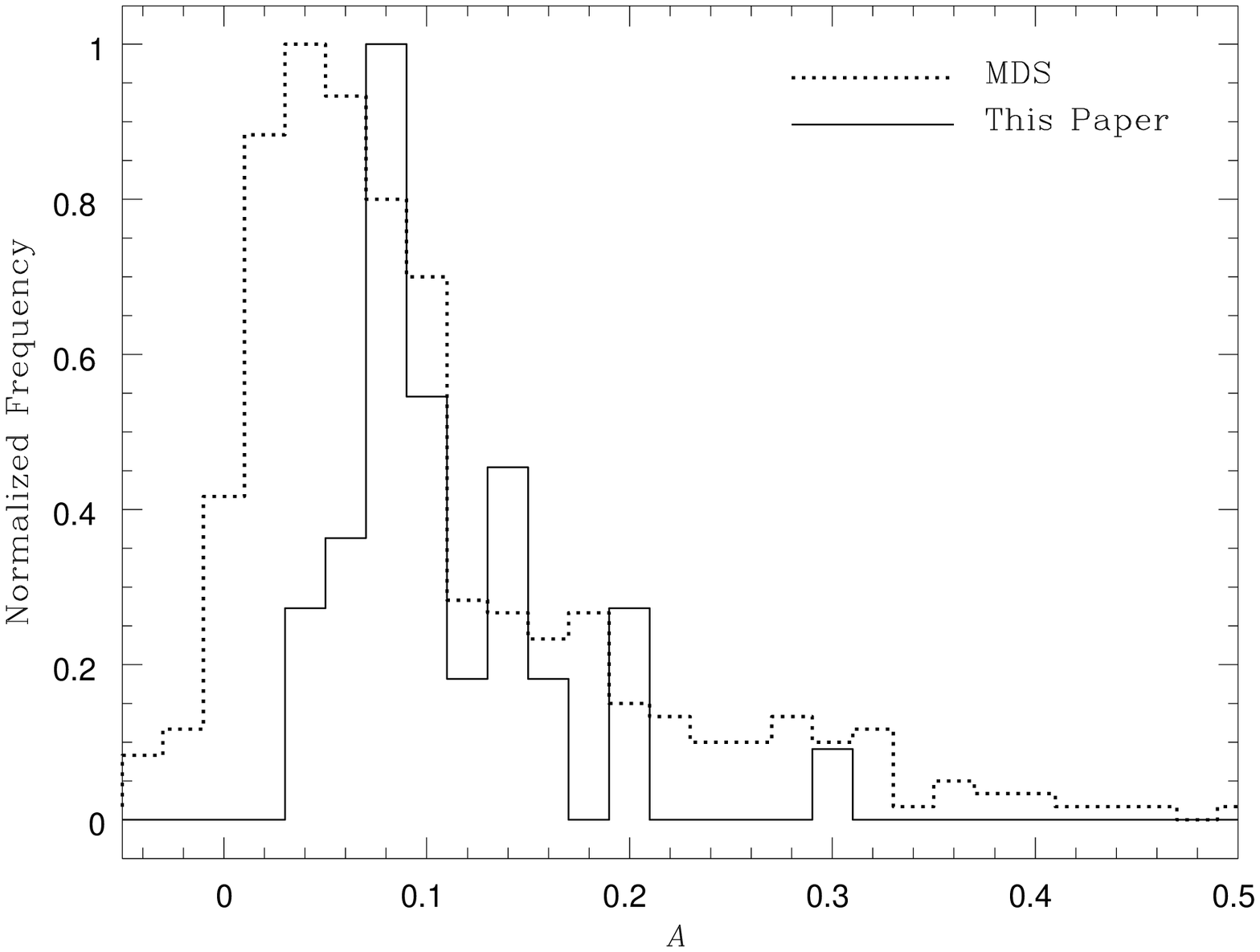}
\caption{The normalized distribution of asymmetries (Abraham's $A$)
for the MDS (dotted line) and our sample (solid line).}
\label{fig:mds}
\end{figure}

Given these caveats, we performed a K--S test to examine whether the
Abraham $A$ asymmetries for the galaxies in our sample are consistent
with having been drawn from the distribution resulting from the MDS
survey.  In Figure~\ref{fig:mds}, we show the normalized distribution
of $A$.  The mode of the asymmetry distribution for our sample
($n=37$) is highly peaked at $A=0.08$ whereas the mode for the MDS
sample ($n=435$) is $A=0.04$. The K--S probability that the two
samples are drawn from the same $A$ distributions is $P_{_{\rm
KS}}=0.0023$, which rules out the null hypothesis of similar
distributions at the 99.77\% confidence level ($2.8~\sigma$).

This is a remarkable result given the uncertainties in the direct
comparison.  The dominant uncertainty is due to our lack of knowledge
of the redshifts of the galaxies in the MDS sample. \citet{abraham96b}
suggest that the extended tail in their $A$ distribution is dominated
by $z>1$ galaxies, which are observed in the rest--frame U--band where
morphological perturbations are significantly pronounced.  Without
redshifts for the MDS galaxies, we cannot select out the $z>1$
galaxies for a more robust test.  However, it is clear that the high
redshift, high asymmetry tail of the MDS galaxy distribution results
in a lower confidence level for the K--S test than if the tail could
be objectively corrected for redshifting effects.

It is important to note that there only two {\MgII} absorption
selected galaxies with $A < 0.04$, below the mode of the MDS
distribution.  This constitutes only 5\% of our sample.  However, 29\%
of the $A$ distribution for the MDS sample is populated below this
value.  It is the difference in this portion of the cumulative
distribution function of $A$ that dominates the K--S statistics and
strongly suggests that {\MgII} absorption selected galaxies tend to
have larger morphological asymmetries than those of the galaxies in
the MDS.  It is quite possible that there is a bias towards galaxies
with slightly larger asymmetries using {\MgII} absorption selection.

\subsection{Interpretation and Context of the Correlation}

Normalizing a galaxy property by the projected distance to the quasar
provides a measure of the influence the galaxy property has out to the
extended regions where the gas is by chance sampled.  If the
influence falls off with distance, then normalizing by $D$ provides a
``scale length correction''. Our results imply that galaxy asymmetry
is an important quantity for predicting the abundance of halo gas and
its covering fraction and that the influence of asymmetry strengthens
with proximity to the galaxy.

For local galaxies there is evidence that morphological perturbations
are associated with a high covering fraction of gas beyond the visible
components of the galaxies \citep[e.g.,][also see Braun
2005]{puche92,swaters97,rand00,fraternali01,fraternali02,barbieri05}.
Our results for intermediate redshifts are consistent with this trend.
Apparently, the greater the morphological perturbations of a galaxy,
the more efficient the process of distributing gas into its halo.
Furthermore, our results suggest that similar absorption strengths can
arise at larger galactocentric distances from galaxies with more
pronounced perturbations or at smaller galactocentric distances from
galaxies with less pronounced perturbations.

Perturbations at the level measured here can be produced via minor
satellite galaxy mergers with the host absorbing galaxy, mild (long
range) galaxy harassment, or continued dynamical interaction with
(i.e., accretion of) the local cosmic web.  Satellite galaxy
interactions can generate Magellanic--type streams, increasing the gas
covering factor out to $\sim 50$ kpc.  These various mechanisms likely
also gently modify the galaxy morphology and induce stochastic periods
of elevated star formation \citep{kaufmann06}. These episodes of star
formation generate galactic fountains, chimneys, superbubbles, and
low--level winds \citep{dahlem97}.  Galaxies exhibiting negligible
morphological perturbations may not have experienced these processes
within the timescale that gas cycles in halos.  These galaxies would
be expected to produce very weak or undetectable {\MgII} absorption
even at moderate impact parameters since they would be deficient in
mechanisms that populate their halos with gas.

This scenario would explain why some bright galaxies close to the
quasar produce weak absorption \citep{cwcweakgals}.  A natural
consequence of this scenario is that it also predicts some galaxies
might have such low quantities of gas in their halos that they would
be excluded from an absorption selected sample even if they lie at
relatively small impact parameter to a quasar.  In the course of our
studies, we discovered a galaxy at $z=0.6720$ in the Q$1317+274$
(TON153) field at an impact parameter of $D=58$~kpc for which the
3~$\sigma$ upper limit on absorption is $W_{r}(2796)=0.007$~{\AA}
\citep{q1317}.  The galaxy asymmetry is $A=0.08$, giving $A/D =
0.0014$. This data point, though an upper limit, is consistent with
expectations from the $A/D$--$W_r(2796)$ correlation, given the
scatter of the data about the maximum likelihood fit. If the
$A/D$--$W_r(2796)$ correlation holds true, it would be expected that
``non--absorbing'' galaxies have very small $A/D$.

The galaxy--absorber pairs for which the absorption profiles are more
than 50\% saturated exhibit a higher level of scatter in the
$A/D$--$W_r(2796)$ correlation than do those that are not highly
saturated.  This may in part be due to the loss of information in
the combined velocity spread and column density distribution of the
gas in the highly saturated profiles.  Our objective examination of
profile saturation resulted in a natural equivalent width break of
$W_r(2796) \simeq 1.4$~{\AA}.  This equivalent width regime may mark a
true transition in the underlying physical mechanisms giving rise to
the gas.  Recently, we have found that galaxies selected by ``weak''
{\MgII} absorption are statistically consistent with those selected by
so--called ``classic'' {\MgII} absorbers
\citep{cwc-china,cwcweakgals}.  For larger $W_r(2796)$, there is a
growing body of evidence that DLAs, wind--driven galaxies, and
strongly interacting galaxies are selected.

Using a statistical argument, \citet{zibetti06} demonstrated that
galaxies selected by {\MgII} absorption with $W_r(2796) \geq
1.1$~{\AA} have colors consistent with star burst galaxies, whereas
galaxies selected by $W_r(2796)<1.1$~{\AA} are consistent with redder,
less active galaxies.  The mean color is consistent with that of an Sb
spiral.  Also using statistical methods, \citet{bouche06} reported an
anti--correlation between the mass of {\MgII} absorbing halos and
$W_r(2796)$.  They argue that this correlation is inconsistent with
virialized gas and suggest that a large fraction of strong {\MgII}
absorbers galaxies [those with $W_r(2796)\gtrsim2$~\AA] are produce by
galactic winds and/or outflows produced by active star formation.
\citet{bond01b} argue and \citet{prochter06} similarly suggest that
{\MgII} absorbers with $W_r(2796) \geq 1.8$~{\AA} arise in galaxies
with elevated star formation, and that the gas is likely wind driven.
\citet{nestor06} imaged the galaxies in the fields of 15 quasars with
``ultra'' strong {\MgII} absorption [$2.7 \leq W_r(2796)\leq 6.0$~\AA]
and found that, compared to a control sample, there is a significant
excess of $L > \hbox{few} \times L_{\ast}$ galaxies and also possibly
interacting galaxies in the quasar fields.

DLAs are also a very different environment than the Lyman--limit and
sub--Lyman limit absorbers in our subsample.  Highly saturated {\MgII}
profiles with moderate velocity spreads preferentially select DLAs
\citep{ellison06}.  It is also observed \citep{rao05} that the
fraction of {\MgII} absorbers that select DLA environments increases
with increasing $W_r(2796)$.  However, it is common for DLAs to
preferentially select smaller impact parameters, which would
systematically elevate their $A/D$ values.

Therefore, we can infer that absorbers with $f_{sat} < 0.5$ arise
predominately from galaxies that have experienced either mild levels
of interactions and/or minor mergers or ongoing accretion of the
cosmic web. These processes would increase the level of star
formation.  The resulting asymmetry may reflect both the gravitational
disturbance of the galaxy as well as surface brightness variations due
to induced star formation. Absorbers with $f_{sat} \geq 0.5$ may arise
predominantly from either DLAs, strong interacting galaxies, or
galaxies with strong winds.  If so, it would not be unexpected that
active star forming galaxies hosting winds might also adhere to the
$A/D$--$W_r(2796)$ correlation, but with a larger scatter due to their
highly saturated absorption profiles.

\section{Concluding Remarks}
\label{sec:conclusion}

We have examined four quantified measures of morphological
asymmetries, $A$, $R_A$, $A_z$, and $D_z$ in a sample of 37 {\MgII}
absorption selected galaxies.  The {\MgII} absorbers were selected
from a database of HIRES and UVES quasar spectra, and from the
literature in a few cases. We modeled the galaxies, observed with
WFPC--2/{\it HST\/} in the F702W and F814W filters, using GIM2D.  The
Abraham $A$ is model independent, whereas $R_A$ is based upon the
residuals of the GIM2D models.  The $A_z$ and $D_z$ are computed
directly from the data, but invoke apertures determined from the GIM2D
models.

We report two main results:

\begin{enumerate}

\item 
We find a statistically significant correlation between the Abraham
$A$ normalized by the quasar galaxy impact parameter, $D$, and the
{\MgII} $\lambda 2796$ rest--frame equivalent width.  The significance
is $3.1~\sigma$ for the our full sample of 37 galaxy--absorber pairs.
However, when we limit the sample to those in which the fraction of
saturated pixes in the {\MgII} $\lambda 2796$ absorption profiles is
$f_{sat} < 0.5$, the $A/D$--$W_r(2796)$ correlation is significant at
the $3.3~\sigma$ level.

\item 
A Kolomorov--Smirnov test comparing the distributions of Abraham $A$
for {\MgII} absorption selected galaxies and magnitude selected
galaxies from the Medium Deep Survey shows the distributions are not
consistent at a confidence level of 99.8\%.  There are caveats related
to the comparison; however, the data show a clear paucity of small $A$
for {\MgII} absorption selected galaxies.  Selecting galaxies by
absorption in quasar spectra may bias against galaxies with negligible
morphological perturbations.

\end{enumerate}

The $A/D$--$W_r(2796)$ correlation is formally statistically
significant, if only at the $3.3~\sigma$ level.  However, it is a
quite remarkable finding given the large variations in kinematic
properties of {\MgII} absorption profiles \citep{cv01}.  As discussed
in \S~\ref{sec:discussion}, the behavior of the correlation is
consistent with a growing body of evidence that the quantity of gas in
galaxy halos is connected to the processes occurring in the galaxies
themselves, and that stronger, highly saturated {\MgII} absorbers are
related to higher levels of activity.

A larger sample of galaxy--absorber pairs with both {\it HST\/}
imaging data and HIRES and/or UVES quasar spectra would be
instrumental for confirming and possibly strengthening the statistical
significance of our results.  One of the major pitfalls of studying
absorption selected galaxies is that it requires intensive work to
obtain a thorough census of the galaxy redshifts in quasar fields.
Thus, we cannot claim that the galaxies used in this study are the
only galaxies associated with the {\MgII} absorbers.

Our results suggest a relationship between the gaseous halo and the
morphological perturbations in the host galaxy normalized to the
impact parameter of the observed absorption. Further insight may be
obtained by comparing the relative kinematics of the absorbing gas
with those of the host galaxy. Only a half--dozen absorption selected
galaxies have been studied at this level of detail
\citep{steidel02,ellison03}.  In addition, further exploration of
correlations between {\CIV} absorption strength and {\MgII} kinematics
\citep{cwc99} with the galaxy morphological perturbations would
provide important clues for better understanding the
$A/D$--$W_{r}(2796)$ correlation.  We are currently working toward
these goals.

\acknowledgments 

We thank Luc Simard for consultation on the details of GIM2D modeling.
We are very grateful to Wal Sargent and Michael Rauch for generously
donating HIRES data for our research.  We also thank Alice Shapley for
expertise with PSF subtraction in WFPC--2 images. We express our
gratitude to the anonymous referee for a careful reading and for
insightful comments that lead to an improved manuscript. Support for
this research through grant HST-AR-10644.01-A was provided by NASA via
the Space Telescope Science Institute, which is operated by the
Association of Universities for Research in Astronomy, Inc., under
NASA contract NAS 5-26555.  G.G.K acknowledges additional support from
Sigma--Xi through the Grants in Aid of Research program.  M.T.M thanks
PPARC for an Advanced Research Fellowship.  J.L.E. acknowledges
support through the New Mexico Space Grant Consortium, which is
administered through NASA.  The imaging data presented in this paper
are based on observations made with the NASA/ESA Hubble Space
Telescope, obtained from the data archive at the Space Telescope
Institute.  STScI is operated by the association of Universities for
Research in Astronomy, Inc. under the NASA contract NAS 5-26555.  Some
spectroscopic data were obtained at the W.M. Keck Observatory, which
is operated as a scientific partnership among the California Institute
of Technology, the University of California and NASA.  The Observatory
was made possible by the generous financial support of the W.M. Keck
Foundation.  Additional spectroscopic data are based on observations
made with European Southern Observatory Very Large Telescope at the
Paranal Observatories under various programs.

{\it Facilities:} \facility{{\it HST\/} (WFPC--2)}, \facility{Keck I
(HIRES)}, \facility{VLT (UVES)}.


\begin{thebibliography}{}

\bibitem[Abraham {\etal}(1996a)]{abraham96a}
Abraham, R. G., Tanvir, N. R., Santiago, B. X., Ellis, R. S.,
Glazebrook, K., \& van den Bergh, S. 1996a, MNRAS, 279, L47

\bibitem[Abraham {\etal}(1996b)]{abraham96b}
Abraham, R. G., van den Bergh, S., Glazebrook, K., Ellis, R. S.,
Santiago, B. X., Surma, P., \& Griffiths, R. E. 1996b, ApJS, 107, 1

\bibitem[Abraham {\etal}(1994)]{abraham94}
Abraham, R. G., Valdes, F., Yee, H. K. C., \& van den Bergh, S. 1994,
ApJ, 432, 75 

\bibitem[Barbieri {\etal}(2005)]{barbieri05} 
Barbieri, C.~V., Fraternali, F., Oosterloo, T., Bertin, G., Boomsma,
R., \& Sancisi, R.\ 2005, \aap, 439, 947

\bibitem[Bergeron \& Boiss\'{e}(1991)]{bb91} 
Bergeron, J., \& Boiss\'{e}, P. 1991, A\&A, 243, 334

\bibitem[Bertin \& Arnouts(1996)]{bertin96}
Bertin, E., \& Arnouts, S. 1996, A\&AS, 117, 393

\bibitem[Bond {\etal}(2001)]{bond01b} 
Bond, N. A., Churchill, C. W., Charlton, J. C., \& Vogt, S. S. 2001,
ApJ, 562, 641 

\bibitem[Bouch{\'e} {\etal}(2006)]{bouche06} 
Bouch{\'e}, N., Murphy, M.~T., P{\'e}roux, C., Csabai, I., \& Wild,
V.\ 2006, \mnras, 371, 495

\bibitem[Braun(2005)]{braun05} 
Braun, R.\ 2005, ASP Conf.~Ser.~331: Extra-Planar Gas, 331,

\bibitem[Chen \& Lanzetta(2003)]{chen03}
Chen, H.-W. \& Lanzetta, K. M. 2003, ApJ, 597, 706

\bibitem[Chen {\etal}(2001)]{chen01} 
Chen, H.-W., Lanzetta, K. M., Webb, J. K., \& Barcons, X. 2001, ApJ,
559, 654

\bibitem[Churchill, Kacprzak, \& Steidel(2005)]{cwc-china}
Churchill, C. W., Kacprzak, G. G., \& Steidel, C. C. 2005, in {\it
Probing Galaxies through Quasar Absorption Lines}, IAU 199
Proceedings, eds.\ P. R. Williams, C.--G. Shu, \& B. M\'{e}nard
(Cambridge: Cambridge University Press), p.\ 24

\bibitem[Churchill {\etal}(2007)]{q1317}
Churchill, C. W., Kacprzak, G. G., Steidel, C. C., \& Evans,
J. L. 2007, ApJ, in press (astro--ph/0612560)

\bibitem[Churchill {\etal}(2006)]{cwcweakgals}
Churchill, C. W., Kacprzak, G. G., Steidel, C. C., \& Murphy,
M. T. 2006, AJ, submitted

\bibitem[Churchill {\etal}(1999a)]{cwc99} Churchill, C.~W., 
Mellon, R.~R., Charlton, J.~C., Jannuzi, B.~T., Kirhakos, S., Steidel,
C.~C., \& Schneider, D.~P.\ 1999a, ApJ, 519, L43

\bibitem[Churchill {\etal}(2000a)]{archiveI}
Churchill, C. W., Mellon, R. R., Charlton, J. C., Jannuzi, B. T.,
Kirhakos, S., Steidel, C. C., \& Schneider, D. P. 2000a, ApJS, 130, 91

\bibitem[Churchill {\etal}(2000b)]{archiveII}
Churchill, C. W., Mellon, R. R., Charlton, J. C., Jannuzi, B. T.,
Kirhakos, S., Steidel, C. C., \& Schneider, D. P. 2000b, ApJ, 543, 577

\bibitem[Churchill {\etal}(1999b)]{weakI}
Churchill, C. W., Rigby, J. R., Charlton, J. C., \& Vogt, S. S. 1999b,
ApJS, 120, 51

\bibitem[Churchill, Steidel, \& Vogt(1996)]{csv96}
Churchill, C. W., Steidel, C. C., \& Vogt, S. S. 1996, ApJ, 471, 164

\bibitem[Churchill \& Vogt(2001)]{cv01}
Churchill, C. W., \& Vogt, S. S. 2001, AJ, 122, 679 

\bibitem[Churchill, Vogt, \& Charlton(2003)]{cvc03}
Churchill, C. W., Vogt, S. S., Charlton, J. C. 2003, AJ, 125, 98

\bibitem[Crampton {\etal}(1995)]{crampton95}
Crampton, D., Le Fevre, O., Lilly, S.~J., \& Hammer, F. 1995, ApJ, 455, 96 

\bibitem[Dahlem(1997)]{dahlem97} 
Dahlem, M.\ 1997, \pasp, 109, 1298

\bibitem[Dekker {\etal}(2000)]{dekker00} 
Dekker, H., D'Odorico, S., Kaufer, A. Delabre, B. \& Kotzlowski
H. 2000, SPIE, 4008, 534

\bibitem[Ellison(2006)]{ellison06}
Ellison, S. L. 2006, MNRAS, 368, 335

\bibitem[Ellison {\etal}(2003)]{ellison03} 
Ellison, S.~L., Mall{\'e}n-Ornelas, G., \& Sawicki, M.\ 2003, ApJ, 589, 709

\bibitem[Evans {\etal}(2007)]{evans07} 
Evans, J. L., {\etal} 2007, ApJ, in prep

\bibitem[Foltz {\etal}(1986)]{foltz86} 
Foltz, C.~B., Weymann, R.~J., Peterson, B.~M., Sun, L., Malkan, M.~A.,
\& Chaffee, F.~H., Jr.\ 1986, \apj, 307, 504

\bibitem[Fraternali {\etal}(2001)]{fraternali01}
Fraternali, F., Oosterloo, T., Sancisi, R., \& van Moorsel, G. 2001,
ApJ, 562, L47

\bibitem[Fraternali {\etal}(2002)]{fraternali02}
Fraternali, F., van Moorsel, G., Sancisi, R., \& Oosterloo, T. 2002,
AJ, 123, 312

\bibitem[Guillemin \& Bergeron(1997)]{gb97}
Guillemin p., \& Bergeron, J. 1997, A\&A, 328, 499

\bibitem[Howell(2004)]{howell04}
Howell D., C. 2004, \\ {\it
http://www.uvm.edu/$\sim$dhowell/StatPages/Resampling/Resampling.html}

\bibitem[Im {\etal}(2002)]{im02}
Im, M., et al. 2002, ApJ, 571, 136

\bibitem[Kacprzak {\etal}(2007)]{glenn-thesis}
Kacprzak, G. G., Churchill, C. W., Steidel, C. C., \& Murphy,
M. T. 2007, ApJS, in preparation

\bibitem[Kacprzak, Churchill, \& Steidel(2005)]{glenn-china}
Kacprzak, G. G., Churchill, C. W., \& Steidel, C. C. 2005, in {\it
Probing Galaxies through Quasar Absorption Lines}, IAU 199
Proceedings, eds.\ P. R. Williams, C.--G. Shu, \& B. M\'{e}nard
(Cambridge: Cambridge University Press), p.\ 80

\bibitem[Kanekar \& Chengalur(2001)]{kanekar01} 
Kanekar, N., \& Chengalur, J.~N.\ 2001, \aap, 369, 42 

\bibitem[Kauffmann {\etal}(2006)]{kaufmann06} 
Kauffmann, G., Heckman, T.~M., De Lucia, G., Brinchmann, J., Charlot,
S., Tremonti, C., White, S.~D.~M., \& Brinkmann, J.\ 2006, \mnras,
367, 1394

\bibitem[Krist \& Hook(2004)]{krist04}
Krist, J. \& Hook, R. 2004, The Tiny Tim User Manual v6.3 (Baltimore:
STScI)

\bibitem[Lanzetta \& Bowen(1992)]{lanzetta92} 
Lanzetta, K.~M., \& Bowen, D.~V.\ 1992, \apj, 391, 48 

\bibitem[Lowenthal {\etal}(1990)]{lowenthal90} 
Lowenthal, J.~D., Caulet, A., Green, R.~F., Hogan, C.~J., Brown,
L.~W., Oliversen, R.~J., \& Woodgate, B.~E.\ 1990, \baas, 22, 805

\bibitem[Marleau \& Simard(1998)]{marleau98} 
Marleau, F.R., \& Simard, L. 1998, ApJ, 505, 585

\bibitem[Miller, Goodrich,  \& Stephens(1987)]{miller87} 
Miller, J.~S., Goodrich,  R.~W., \& Stephens, S.~A.\ 1987, \aj, 94, 633 

\bibitem[Monet {\etal}(1998)]{usno2}
Monet, D., {\etal} 1998, USNO--SA2.0: A Catalog of Astrometric
Standards (Washington: US Nav. Obs.)

\bibitem[Murphy(2006)]{popler}
Murphy, M. T. 2006, {\sc uves popler}, \\ {\it
http://www.ast.cam.ac.uk/$\sim$mim/UVES\_popler.html}

\bibitem[Nestor {\etal}(2006)]{nestor06}
Nestor, D.~B., Turnshek, D.~A., Rao, S.~M., \& Quider, A.~M.  2006, ApJ, in press
(astro--ph/0610760)

\bibitem[Prochter {\etal}(2006)]{prochter06} 
Prochter, G.~E., Prochaska, J.~X., \& Burles, S.~M.\ 2006, \apj, 639,
766

\bibitem[Puche {\etal}(1992)]{puche92}
Puche, D., Westpfahl, D., Brinks, E., \& Roy, J. 1992, AJ, 103, 1841

\bibitem[Rand(2000)]{rand00} 
Rand, R. 2000, ApJ, 537, 13                 

\bibitem[Rao(2005)]{rao05} 
Rao, S.~M.\ 2005, in {\it Probing Galaxies through Quasar Absorption
Lines}, IAU 199 Proceedings, eds.\ P. R. Williams, C.--G. Shu, \&
B. M\'{e}nard (Cambridge: Cambridge University Press), p.\ 125

\bibitem[Rao \& Turnshek(2000)]{rao00}
Rao, S.~M., \& Turnshek, D.~A.\ 2000, \apjs, 130, 1 

\bibitem[Sargent, Boksenberg \& Steidel(1988)]{sargent88} 
Sargent, W.~L.~W., Boksenberg, A., \& Steidel, C.~C.\ 1988, \apjs, 68,
539

\bibitem[Savage, Tripp, \& Lu(1998)]{savage98}
Savage, B. D., Tripp, T. M. \& Lu, L. 1998, AJ, 115, 436

\bibitem[Schade {\etal}(1995)]{schade95}
Schade. D., Lilly, S. J., Crampton, D., Hammer, F., Le Fevre, O., \&
Tresse, L. 1995, ApJ, 451, L1

\bibitem[Simard {\etal}(1999)]{simard99}
Simard, L., {\etal} 1999, ApJ, 519, 563

\bibitem[Simard {\etal}(2002)]{simard02}
Simard, L., Willmer, C. N. A., Vogt, N. P., Sarajedini, V. L.,
Philips, A. C., Weiner, B. J., Koo, D. C., Im, M., Illingworth, G. D.,
\& Faber, S. M. 2002, ApJS, 142, 1

\bibitem[Spinrad {\etal}(1993)]{spinrad93}
Spinrad, H., Filippenko, A. V., Yee, H. K., Ellingson, E., Blades,
J. C., Bahcall, J. N.; Jannuzi, B. T.; Bechtold, J., \& Dobrzycki,
A. 1993 AJ, 106, 1

\bibitem[Steidel(1995)]{steidel95} 
Steidel, C. C. 1995, in QSO Absorption Lines, ed.\ G. Meylan,
(Springer--verlag: Berlin Heidelberg), p.\ 139

\bibitem[Steidel(1998)]{steidel98} 
Steidel, C. C. 1998, in Galactic Halos: A UC Santa Cruz Workshop, ed.\
D. Zaritsky, ASP Conference Proceeding, V136, p167

\bibitem[Steidel {\etal}(1997)]{steidel97} 
Steidel, C. C., Dickinson, M., Meyer, D. M., Adelberger, K. L., \&
Sembach, K. R. 1997, ApJ, 480, 586

\bibitem[Steidel, Dickinson, \& Persson(1994)]{sdp94}
Steidel, C. C., Dickinson, M., \& Persson, S. E.  ApJ, 437, L75

\bibitem[Steidel {\etal}(2002)]{steidel02} 
Steidel, C. C., Kollmeier, J. A., Shapely, A. E., Churchill, C. W.,
Dickinson, M., \& Pettini, M. 2002, ApJ, 570, 526

\bibitem[Steidel \& Sargent(1992)]{steidel92} 
Steidel, C.~C., \& Sargent, W.~L.~W.\ 1992, \apjs, 80, 1 

\bibitem[Stetson(1989)]{stetson89}
Stetson, P. B. 1989, Advanced School of Astrophysics, (Univerisidade
de Sao Paulo), p1

\bibitem[Swaters, Sancisi \& van der Hulst(1997)]{swaters97}
Swaters, R. A., Sancisi, R., \& van der Hulst, J. M. 1997, ApJ, 491, 140

\bibitem[Turnshek {\etal}(1989)]{turnshek89} 
Turnshek, D.~A., Wolfe, A.~M., Lanzetta, K.~M., Briggs, F.~H., Cohen,
R.~D., Foltz, C.~B., Smith, H.~E., \& Wilkes, B.~J.\ 1989, \apj, 344,
567

\bibitem[Vogt {\etal}(1994)]{vogt94}
Vogt, S. S., {\etal} 1994, SPIE, 2198, 362

\bibitem[Whyte {\etal}(2002)]{whyte02}
Whyte, L. F., Abraham, R. G., Merrifield, M. R., Eskridge, P. B.,
Frogel, J. A., \& Pogge, R. W. 2002, MNRAS, 336, 1281

\bibitem[Whitmore(1995)]{whitmore95} 
Whitmore, B. 1995, in Calibrating {\it Hubble Space Telescope}: Post
Servicing Mission, eds.\ Koratkar A. \& Leitherer, C., (Baltimore:
STScI)

\bibitem[Zheng {\etal}(2005)]{zheng05} 
Zheng, X.Z., Hammer, F., Flores, H., Ass\'{e}mat, F., \& Rawat,
A. 2005, A\&A, 435, 507

\bibitem[Zibetti {\etal}(2006)]{zibetti06} 
Zibetti, S., Menard, B., Nestor, D.~B., Quider, A.~M., Rao, S.~M., 
\& Turnshek, D.~A.\ 2006, ApJ, in press (astro--ph/0609760) 


\bibitem[Zibetti {\etal}(2005)]{zibetti05} 
Zibetti, S., M{\'e}nard, B., Nestor, D., \& Turnshek, D.\ 2005, ApJ,
631, L105

\end{thebibliography}
\end{document}